\documentclass[aps,twocolumn,prb,superscriptaddress]{revtex4-2}
\usepackage{amsfonts, amssymb, amsmath,amsthm,graphicx}
\usepackage{mathrsfs}
\usepackage{subfigure}
\usepackage{multirow}
\usepackage{dcolumn}
\usepackage{bm}
\usepackage{color}
\usepackage{dsfont}
\usepackage[usenames,dvipsnames]{xcolor}

\allowdisplaybreaks

\newcommand{\E}{\mathcal{E}}

\renewcommand{\bf}[1]{\textnormal{\textbf{#1}}}
\newcommand{\BZ}{\textnormal{\text{BZ}}}
\newcommand{\tr}{\textnormal{\text{Tr}}}

\newcommand{\Gr}{\textnormal{\text{Gr}}}

\newcommand{\ket}[1]{| #1 \rangle}
\newcommand{\bra}[1]{\langle #1|}

\newcommand{\bk}{\mathbf{k}}

\newcommand{\eps}{\epsilon}

\newcommand{\dki}{\partial_i}
\newcommand{\dkj}{\partial_j}

\newcommand{\cA}{\mathcal{A}}
\newcommand{\cJ}{\mathcal{J}}

\renewcommand{\Im}{\ensuremath{\mathrm{Im}\,}}
\renewcommand{\Re}{\ensuremath{\mathrm{Re}\,}}

\usepackage[colorlinks=true,allcolors=blue]{hyperref}

\begin{document}
\title{Nontrivial quantum geometry of degenerate flat bands}
\author{Bruno Mera}
\affiliation{Advanced Institute for Materials Research (WPI-AIMR), Tohoku University, Sendai 980-8577, Japan}
\author{Johannes Mitscherling}
\affiliation{Max Planck Institute for Solid State Research, Heisenbergstrasse
1, 70569, Stuttgart, Germany}
\affiliation{Department of Physics, University of California, Berkeley, California 94720, USA}

\date{\today}

\begin{abstract}
    The importance of the quantum metric in flat-band systems has been noticed recently in many contexts such as the superfluid stiffness, the dc electrical conductivity, and ideal Chern insulators. Both the quantum metric of degenerate and nondegenerate bands can be naturally described via the geometry of different Grassmannian manifolds, specific to the band degeneracies. Contrary to the (Abelian) Berry curvature, the quantum metric of a degenerate band resulting from the collapse of a collection of bands is not simply the sum of the individual quantum metrics. We provide a physical interpretation of this phenomenon in terms of transition dipole matrix elements between two bands. By considering a toy model, we show that the quantum metric gets enhanced, reduced, or remains unaffected depending on which bands collapse. The dc longitudinal conductivity and the superfluid stiffness are known to be proportional to the quantum metric for flat-band systems, which makes them suitable candidates for the observation of this phenomenon.
\end{abstract}

\maketitle
{\it Introduction.---}
The quantum metric provides a measure of distance between wave functions in the study of phase transitions~\cite{Zanardi2006, Rezakhani:10, Carollo2020} and is crucial in the modern theory of polarization~\cite{Resta2011} due to its relation to the size of maximally localized Wannier functions~\cite{Marzari1997, Marzari2012}. 
Recently, nontrivial relations between the quantum metric and the Berry curvature have been understood via the underlying K\"{a}hler geometry of the space of quantum states~\cite{Ozawa2021, Mera2021a, Mera2021b} 
and have been successfully applied to ideal Chern bands and fractional Chern insulators~\cite{Varjas2021, Wang2021a, Wang2021b, Ledwith2021, Northe2021, Parker2021}. In materials with highly quenched band width, the quantum metric yields the dominant contribution to the superfluid stiffness~\cite{Peotta2015, Hofmann:20, Torma2021, Kitamura2021a, Rossi2021, Verma2021,  Lewandowski:21, Huhtinen2022, Lau2022, Hofmann:22} and the dc electrical conductivity~\cite{Mitscherling2022, Bouzerar:22}. Here, inequalities for the quantum metric related to the Chern number~\cite{Roy2014, Ozawa2021, Mera2021a, Wang2021a}, the Euler characteristic \cite{Xie2020}, or obstructed Wannier functions~\cite{Herzog2022} result in lower bounds with direct implications for moir\'e materials such as twisted bilayer graphene~\cite{Hu2019, Julku2020, Xie2020} and untwisted heterostructures with flat bands such as rhombohedral trilayer graphene~\cite{Mitscherling2022}. Especially in the last years, connecting the newly identified importance of the quantum metric in many fields with new insights on fundamental properties of the quantum metric has been established as a powerful research direction~\cite{Souza2008, Matsuura2010, Ma2010, Neupert2013, Kolodrubetz2013, Gao2014, Srivastava2015, Albert2016, Piechon2016, Freimuth2017, Kolodrubetz2017, Iskin2018, Ozawa2018, Gao2019, Zhang2019, Hu2020, Ma2020, Rattacaso2020, Zhao2020, Ahn2021, Bhalla2021, Cayssol2021, Graf2021, Hauser2021, Hwang2021, Julku2021, Kitamura2021, Leblanc2021, Penner2021, Wang2021, Xiao2021, Zhang2021, Abouelkomsan2022, Lin2022}.

Geometric quantities such as the quantum metric arise naturally in the description of interband effects in multiband systems. Interband transitions are described by the product of two Berry connection coefficients, defining transition dipole moments.
In transport, such transitions can be induced, for instance, by finite frequencies of the external electric fields~\cite{Gao2020, Holder2020, Chen2021, Liang2021, Ma2021, Watanabe2021, Ahn2022, Li2022} or virtual band excitations~\cite{Mitscherling2020, Mitscherling2022, Pickem2022}. Similar contributions are also found for nonuniform electric fields~\cite{Lapa2019, Kozii2021} and in spectroscopy~\cite{Ozawa2019, Klees2020, Klees2021, Topp2021, Gersdorff2021, Chen2022}. For two-band systems, the symmetric and antisymmetric parts of the transition dipole moment are proportional to the quantum metric and the Berry curvature, respectively. Since every transition between pairs of bands might be weighted differently, for instance, due to different band occupations, such an identification is possible only in special situations for more than two bands~\cite{Mitscherling2022}, which make general multiband systems promising candidates for new quantum geometric phenomena. 

In this paper, we analyze the quantum geometry of degenerate bands by using their relation to the geometry of Grassmannians. It has been noticed before by Peotta and T\"orm\"a \cite{Peotta2015} that the quantum metric 
is not additive upon collapse of a collection of bands. However, the physical implications have not been investigated so far. The recently discovered quantum metric contribution to the dc electrical conductivity~\cite{Mitscherling2022} provides a simple theory, which yields a physical quantity proportional to the integrated quantum metric for flat bands and captures the crossover between nondegenerate and (effectively) degenerate bands. For a flat-band toy model, we show that the quantum metric gets enhanced, reduced, or remains unaffected, due to the nontrivial quantum metric of the collapsed bands and, as a consequence, the dc longitudinal conductivity, which we calculate following Ref.~\cite{Mitscherling2022}, exhibits the same behavior. Our results are directly applicable to all the physical observables related to the quantum metric, such as the superfluid stiffness. 

{\it The Bloch bundle.---}
We give a self-contained review of the differential geometry of band theory (see also the Supplemental Material (SM)~\cite{SupplMat}), which is the framework we use.
The profitable relation between geometry and quantum mechanics has already been used in different contexts~\cite{Provost1980, Wilczek1984, Anandan1990, Ashtekar1997, Ozawa2021, Mera2021a, Mera2021b, Alvarez2020, Gonzalez2020, Holder2021, Ahn2022, Smith2022}. Under the assumption of short-range hopping amplitudes, a tight-binding Hamiltonian with $N$ internal degrees of freedom,
\begin{align}
    \label{eqn:H}
    H=\sum_\bk\sum_{i,j=1}^{N}\Psi^\dagger_{i,\bk} \,H^{}_{ij}(\bk) \,\Psi^{}_{j,\bk},
\end{align}
gives rise to an $N\times N$ Hermitian matrix $H(\bk)=[H_{ij}(\bk)]_{1\leq i,j\leq N}$, which smoothly depends on momentum $\bk\in\BZ^d$ over the $d$-dimensional Brillouin zone $\BZ^d$. Here, $\Psi^\dag_{i,\bf{k}}$ and $\Psi^{}_{i,\bf{k}}$ are fermionic creation and annihilation operators at $\bf{k}$ and internal degree of freedom $i$, respectively. For fixed $\bk$, the Hermitian matrix $H(\bk)$ acts on the vector spaces of Bloch wave functions denoted by $\E_{\bk}$. The collection of all these vector space forms the \emph{Bloch (vector) bundle} $\E\overset{\pi}{\longrightarrow} \BZ^d$. The bundle $\E$ comes equipped with a connection $\nabla$---known as the \emph{Berry connection}. It 
is related to the position operator in the Bloch representation by $\bf{r}=i\nabla$. In the global gauge of $\mathcal{E}$ provided by $\Psi_{i,\bf{k}}^\dagger\ket{0}$, $i=1,\dots,N$, $\nabla$ is simply the exterior derivative $d=\sum_{j=1}^{d}dk_j\frac{\partial}{\partial k_j}$. Since $d^2=0$, this connection is \emph{flat}, i.e., it has no \emph{curvature}, which is consistent with the fact that position operators commute. 

The Hermitian matrix $H(\bk)$ is diagonalized by the unitary matrix $U(\bk)=[\ket{u_{1,\bk}},\dots, \ket{u_{N,\bk}}]$ involving Bloch wave functions $\ket{u_{m,\bk}}$ as columns. Whereas $H(\bk)$ and its spectrum, i.e., the energy bands $E_m(\bk)$, are smooth and globally defined, $U(\bk)$ does not need to be smoothly defined globally. In fact, at each momentum, it is defined up to multiplication on the right by a unitary matrix preserving the diagonal matrix of eigenvalues of $H(\bk)$. Thus, the Bloch Hamiltonian induces a splitting of the vector space $\E_{\bk}\cong \mathbb{C}^N$ into mutually orthogonal vector subspaces with dimensions given by the degeneracies of the eigenvalues at that point $\bk\in \BZ^d$. Provided the eigenvalues do not cross, these decompositions glue together and provide a splitting of the Bloch bundle $\E$ into vector subbundles of ranks given by the degeneracies of the bands. If we write the Berry connection $\nabla$ on $\E$ using the local frame field provided by $U(\bk)$, we find non-trivial local connection coefficients, i.e., a (local) gauge field 
\begin{align}
\label{eqn:BerryConnection}
A(\bk)=U(\bk)^{-1}dU(\bk)=[\bra{u_{m,\bk}}d\ket{u_{n,\bk}}]_{1\leq m,n\leq N} \, .
\end{align}
The quantity $A$ is the pullback of the Maurer-Cartan $1$-form of $\text{U}(N)$ under the locally defined map $\bk\mapsto U(\bk)$. The non-vanishing of $A$ does not violate the flatness of the connection on $\E$, since $dA+A\wedge A=0$.

{\it Insulators. - } For band insulators, the ground state is obtained by filling the entire bands below the Fermi level $E_{F}$. The \emph{Fermi projector} associated with these occupied bands $P_{F}(\bk)=\Theta(E_F -H(\bk))$, with $\Theta$ the Heaviside step function, provides a splitting of the Bloch bundle as
\begin{align}
\E=\text{Im}(P_{F})\oplus \text{Ker}(P_{F})=\text{Im}(P_{F})\oplus \text{Im}(Q_{F}),
\end{align}
where $Q_{F}(\bk)=I_{N}-P_{F}(\bk)$ with identity matrix $I_N$. The \emph{occupied Bloch bundle} $\text{Im}(P_{F})$ is the vector subbundle of $\E$ whose fiber at $\bk$ is the image  $\text{Im}(P_{F}(\bk))$. $\text{Im}(Q_{F})$ and $\text{Ker}(P_{F})$ are defined similarly. Although $\mathcal{E}$ is a trivial vector bundle, the subbundles $\text{Im}(P_{F})$ and $\text{Im}(Q_{F})$ are not necessarily trivial, leading to rich topological effects such as the quantum anomalous Hall effect~\cite{Thouless1982, Niu1985}.
The Fermi projector defines a map $P_{F}:\BZ^d \to \text{Gr}_{N_{occ}}(\mathbb{C}^N)$, where $\text{Gr}_{N_{occ}}(\mathbb{C}^N)=\text{U}(N)/\big(\text{U}(N_{occ})\!\times\! \text{U}(N\!-\!N_{occ})\big)$ denotes the \emph{Grassmannian of }$N_{occ}$-\emph{dimensional subspaces of} $\mathbb{C}^N$ with $N_{occ}$ being the number of bands below $E_F$. 

{\it Berry curvature and quantum metric.---} 
For a smooth orthogonal projector $P:\BZ^d\to\text{Gr}_{r}(\mathbb{C}^N)$ of some rank $r$, the Berry connection $\nabla$ on $\mathcal{E}$ does not necessarily preserve the sections of $\text{Im}(P)$ because the components of Eq.~\eqref{eqn:BerryConnection}, for $\ket{u_{n,\bf{k}}}$ taking values in $\text{Im}(P)$ and $\ket{u_{m,\bf{k}}}$ in $\text{Im}(Q)$, 
can be nontrivial. The composition $P\nabla$, acting on sections of $\text{Im}(P)\subset \mathcal{E}$, defines the projected Berry connection which, in general, is no longer flat. Its curvature, known as the \emph{Berry curvature}, is the $2-$form~\cite{SupplMat}
\begin{align}
\Omega=(P\nabla) \wedge (P\nabla)=PdP\wedge dP P \, .
\label{eq: Berry curvature}
\end{align}
The \emph{Abelian Berry curvature} is $F=\tr \left(\Omega\right)$, where the trace is taken over the internal indices.

We obtain further insights and properties by exploring the role of the map $P$ to the relevant Grassmannian.  If one recalls the definition of the Fubini-Study K\"{a}hler form $\omega_{FS}$ on the Grassmannian---a K\"{a}hler manifold~\cite{Cannas2008}--- one finds that $F$ equals the pullback under $P$ of $2i \omega_{FS}$ \cite{Mera2021a, Mera2021b},
\begin{align}
F=2i P^*\omega_{FS}=\tr\left( PdP\wedge dP\right)
\, .
\label{eq: Abelian Berry curvature}
\end{align}
Furthermore, 
the pullback of the Fubini-Study metric $g_{FS}$ of the Grassmannian defines the \emph{quantum metric},
\begin{align}
g=P^*g_{FS}=\tr\left( PdPdP\right)=\frac{1}{2}\tr\left( dP dP\right) \, .
\label{eq: quantum metric definition}
\end{align}
Using the Cauchy-Schwarz inequality associated with the Hermitian form $g_{FS}+i\omega_{FS}$ of the Grassmannian, it follows that
$g^{ii}(\bk)g^{jj}(\bk)-g^{ij}(\bk)g^{ij}(\bk)\geq
\left|F^{ij}(\bk)/2\right|^2\!\!\!$ for $i,j\in\{1,\dots,d\}$ 
with $g\!=\!\sum_{i,j}g^{ij} dk_idk_j$
and $F=(1/2)\sum_{i,j}F^{ij} dk_i\wedge dk_j$ \cite{Mera2021a}. 
This identity implies an inequality between the Chern number and the quantum volume \cite{Ozawa2021,Mera2021a,Mera2021b,Mera2022} and
\begin{align}
    g^{ii}(\bk)+g^{jj}(\bk)\geq |F^{ij}(\bk)|\,,
    \label{eqn:inequality}
\end{align} 
which generalizes the result known for two \cite{Roy2014} to $d$ dimensions. Equation \eqref{eqn:inequality} has been used to identify lower bounds on quantities involving the quantum metric \cite{Peotta2015, Mitscherling2022}. 

{\it Isolated bands.---} 
The previous results can be directly applied to other relevant projectors. When an energy band $n$ is isolated, i.e., it does not cross any other band, there is a well-defined orthogonal projector $P_n(\bk)$ at each $\bk\in\BZ^d$ with fixed rank $N_{n}\in\{1,\dots,N\}$, which corresponds to the band degeneracy. $P_n(\bk)$
defines a map $P_n:\BZ^d\to\Gr_{N_n}(\mathbb{C}^N)$. For a \emph{nondegenerate} band $N_n=1$, $P_n$ assigns the ray associated with the corresponding Bloch wave-function $\ket{u_{n,\bk}}$ of $H(\bk)$ to each $\bk\in\BZ^d$. We have $\Gr_{1}(\mathbb{C}^N)\cong\mathbb{C}P^{N-1}$, which 
is commonly known as 
{\it Bloch sphere} for a two-band system. For an $N_n-$fold degenerate band, the map $P_n$ gives rise to an associated vector bundle $\text{Im}(P_n)\overset{\pi}{\longrightarrow}\BZ^d$ whose fibers are spanned by an orthonormal basis of corresponding $N_n$ eigenfunctions
$\ket{u_{ns,\bk}}$, $s=1,...,N_n$.
Using Eqs.~\eqref{eq: quantum metric definition} and \eqref{eq: Abelian Berry curvature}, the explicit formulas for the quantum metric and the Abelian Berry curvature of band $n$
are \cite{SupplMat}
\begin{align}
g_n(\bk)
&\!=\!\sum_{s=1}^{N_n} \sum_{i,j=1}^{d}\!\bra{\dki u_{ns,\bk}}Q_n(\bk)\ket{\dkj u_{ns,\bk}}dk_idk_j \,,
\label{eq: quantum metric} \\
F_n(\bk)
&\!=\!
\sum_{s=1}^{N_n} \sum_{i,j=1}^{d}\!\bra{\dki u_{ns,\bk}}Q_n(\bk)\ket{\dkj u_{ns,\bk}}dk_i\wedge dk_j \,,
\label{eq: symplectic form}
\end{align}
where $Q_n(\bk)=I_{N}-P_n(\bk)$ and $\partial_i\equiv \partial/\partial k_i$, $i=1,\dots,d$.

{\it Nonadditivity of the quantum metric.---} Let us consider a \emph{split} band projector, i.e., an orthogonal projector $P_n$ of rank $N_{n}$ which decomposes into the sum of mutually orthogonal projectors
\begin{align}
P_n(\bf{k})=P_1(\bf{k})+P_2(\bf{k}),
\end{align}
with $P_1$ and $P_2$ having ranks $N_1$ and $N_2$, respectively.
This situation occurs when two bands described by $P_1$ and $P_2$ (effectively) degenerate into one by tuning some external parameter. 
The main result that we want to emphasize, previously noted in~\cite{Peotta2015}, is that the quantum metric $g_n$ of a split band is not generally the sum of the quantum metrics $g_1,g_2$ of each of the individual bands. Instead,
\begin{align}
g_n=g_1+g_2 +\tr\left(dP_1dP_2\right).
\label{eqn:nonadditivity}
\end{align}
This additional term can even render $g_n=0$ if $P_1+P_2$ is a constant projector, i.e., if $\text{Im}(P_n)$ is a trivial bundle 
\cite{SupplMat}.
In contrast, the Abelian Berry curvature $F_n$ of the split band
is equal to the sum of the Abelian Berry curvatures of each band.
The upper results can be easily generalized to multiply split bands.

We now give a physical interpretation of the nonadditivity property. If we write $P_i(\bf{k})=\sum_{m=1}^{N_i}\ket{u_{im,\bf{k}}}\bra{u_{im,\bf{k}}}$, $i=1,2$, then the mixed term $\tr\left(dP_1dP_2\right)$ can be written as
\begin{align}
\tr\left(dP_1dP_2\right)=-2\sum_{s=1}^{N_1}\sum_{l=1}^{N_2}|\bra{u_{2l,\bf{k}}}d\ket{u_{1s,\bf{k}}}|^2,    
\end{align}
which is the sum of the squares of all possible transition dipole matrix elements $i\bra{u_{2l,\bf{k}}}\partial_j\ket{u_{1s,\bf{k}}}$, $s=1,\dots, N_1$, $l=1,\dots, N_2$, $j=1,\dots, d$, between the two bands. The nonvanishing of this contribution tells us that, if $P_1(\bf{k})$ and $P_{2}(\bf{k})$ described isolated degenerate bands separated by some gap, then states can be excited from one band to another induced, for instance, by finite frequencies of the external electric fields or virtual band excitations.

{\it DC electrical conductivity.---} 
We apply the general results presented above to the dc electrical conductivity tensor $\sigma^{ij}$, which relates the current and the external electric field via $\cJ^i = \sum_{j=1}^{d} \sigma^{ij} E^j$. The conductivity tensor can be conveniently decomposed into $\sigma^{ij}=\sigma^{ij}_\text{intra}+\sigma^{ij,s}_\text{inter}+\sigma^{ij,a}_\text{inter}$ \cite{Mitscherling2020}. In the following, we focus on the (symmetric) quantum metric contribution 
\cite{Mitscherling2020, Mitscherling2022}, 
\begin{align}
    \sigma^{ij,s}_\text{inter}= \frac{e^2}{\hbar}\!\!\int\!\!\!\frac{d^d\bk}{(2\pi)^d} \sum^N_{\underset{n\neq m}{n,m=1}} w^{\text{inter},s}_{nm}(\bk)\,\, g^{ij}_{nm}(\bk) \, , \label{eqn:sigmaS}
\end{align}
with electric charge $e$, reduced Planck's constant $\hbar$, and summation over pairs of the $N$ bands. We have $g^{ij}_{nm}(\bk)\equiv\Re\big[r^i_{nm}(\bk)\,r^j_{mn}(\bk)\big]$ involving the transition dipole matrix element $r^{i}_{nm}\equiv i\bra{u_{n,\bf{k}}}\partial_i\ket{u_{m,\bf{k}}}$ (cf. Eq.~\eqref{eqn:BerryConnection}). 
Each transition is weighted by $w^{\text{inter},s}_{nm}(\bk)\equiv-\pi(E_{n,\bk}\!-\!E_{m,\bk})^2\!\!\int\!\!d\eps f'(\eps)\cA_n(\bk,\eps)\cA_m(\bk,\eps)$, where $\cA_n(\bk,\eps)= \Gamma/\big(\pi \big[\Gamma^2+(\eps+\mu-E_{n,\bk})^2\big]\big)^{-1}$ is the spectral function of band $n$ with chemical potential $\mu$ and phenomenological relaxation rate $\Gamma$. $f(\eps)=\big[\exp{(\eps/k_B T)}+1\big]^{-1}$ is the Fermi function with Boltzmann constant $k_B$ and temperature $T$. We present the analogous results for the intraband and the (antisymmetric) Berry curvature contribution $\sigma^{ij}_\text{intra}$ and $\sigma^{ij,a}_\text{inter}$ in the SM~\cite{SupplMat}.

{\it Conductivity of degenerate bands.---} 
We consider $r$ isolated bands. Each band $n$ is $N_n$-fold degenerate with $E_{n,\bk}\equiv E_{(ns),\bk}$, where $s=1,...,N_n$. 
We notice that $w^{\text{inter},s}_{nm}\equiv w^{\text{inter},s}_{(ns)(ml)}$ only depends on the degenerate eigenvalues
and are, thus, equal for all $s=1,...,N_n$ and $l=1,...,N_m$. In particular, interband transitions within a degenerate band vanish. Using this, we equivalently write the formula in Eq.~\eqref{eqn:sigmaS} as 
\begin{align}
 &\sigma^{ij,s}_\text{inter}=\frac{e^2}{\hbar}\!\!\int\!\!\!\frac{d^d\bk}{(2\pi)^d}\sum^r_{\underset{n\neq m}{n,m=1}} 
 w^{\text{inter},s}_{nm}(\bk)\,\, \widehat{ g}^{\,ij}_{nm}(\bk) \, , \label{eqn:sigmaSdeg}
\end{align}
with summation only over pairs of the $r$ different degenerate subspaces and 
\begin{align}
   \label{eqn:gnm} \widehat g^{\,ij}_{nm}(\bk)
   &\equiv \sum_{s=1}^{N_n}\sum_{l=1}^{N_m} \Re\!\Big[i\langle u_{ns,\bk}|\dki u_{ml,\bk}\rangle i\langle u_{ml,\bk}|\dkj u_{ns,\bk}\rangle\Big] \, ,
\end{align}
which includes the remaining summation within the two involved degenerate subspaces. 
We prove that $\widehat{ g}^{\,ij}_{nm}$ is invariant under $\text{U}(N_n)\times \text{U}(N_m)$-gauge transformations \cite{SupplMat},
which shows the gauge-invariance of the conductivity in Eq.~\eqref{eqn:sigmaS} and each term in Eq.~\eqref{eqn:sigmaSdeg}. 

As a first application, we study a system composed of two independent copies of a single system with Bloch Hamiltonian $H(\bk)$, with $N$ nondegenerate bands. 
Then, the eigenvalues $E_{n,\bk}$ of the Hamiltonian $H'(\bk)\equiv H(\bk)\oplus H(\bk)$ are two-fold degenerate with eigenvectors $|u_{n1,\bk}\rangle=(|u_{n,\bk}\rangle,0)$ and $|u_{n2,\bk}\rangle=(0,|u_{n,\bk}\rangle)$, where $\ket{u_{n,\bk}}$ is the corresponding eigenvector of $H(\bk)$. From Eq.~\eqref{eqn:gnm} it follows that $\widehat {g}^{\,'ij}_{nm}=2\,\widehat{g}^{\,ij}_{nm}$, which is the expected trivial enhancement.
Whereas the intraband contribution $\sigma^{ij}_\text{intra}$ of an $N_n$-degenerate band $n$ is always enhanced by a factor $N_n$ in relation to the non-degenerate case~\cite{SupplMat}, this is, however, not generally true for the quantum metric contribution in Eq.~\eqref{eqn:sigmaSdeg} as we will see in the following.

{\it Underlying Grassmannian geometry.---} 
Using Eqs.~\eqref{eq: quantum metric} and \eqref{eqn:gnm}, the relation between 
$\widehat g^{\,ij}_{nm}$ involving a specific degenerate band $n$ and the quantum metric components $g^{ij}_n$ 
induced by the projection $P_n(\bk)=\sum^{N_n}_{s=1}|u_{ns,\bk}\rangle\langle u_{ns,\bk}|$ onto this band is   
\begin{align}
    \sum_{\underset{m\neq n}{m=1}}^r \widehat g^{\,ij}_{nm}(\bk)=g^{ij}_{n}(\bf{k}) \, .
    \label{eqn:relation}
\end{align}
This shows the close relation between the gauge-invariant transition dipole moments defined in Eq.~\eqref{eqn:gnm} 
involving an $N_n$-fold degenerate band and the geometry of the corresponding Grassmannian $\Gr_{N_n}(\mathbb{C}^N)$. 

The conductivity in Eq.~\eqref{eqn:sigmaSdeg} and the identity \eqref{eqn:relation} differ by the transition-dependent weights $w^{\text{inter},s}_{nm}$. These weights drastically simplify for a clean metal and in flat-band systems. In presence of a $(d-1)$-dimensional Fermi surface, we have  
\begin{align}
    \sigma^{ij,s}_\text{inter}= -\frac{2\Gamma e^2}{\hbar}\sum^r_{n=1} \int\!\!\!\frac{d^d\bk}{(2\pi)^d}\,\,f'(E_{n,\bk}-\mu)\,\,g^{ij}_{n}(\bk) \, ,
    \label{eqn:sigmaSclean}
\end{align}
if the band gaps are small on the scale of $\Gamma$ and the metric is almost constant on the momentum scale, in which the variation of the dispersion is of order $\Gamma$ \cite{Mitscherling2018, Mitscherling2020, Mitscherling2022}. We see that each band contribution involves the quantum metric that corresponds to the underlying Grassmannian.
Since the intraband contribution scales as $1/\Gamma$ in the clean limit, significant corrections due to the quantum metric are expected only for small band gaps $\Delta\sim \Gamma$, for instance, at the onset of order at quantum critical points \cite{Mitscherling2018, Bonetti2020}. Let us assume an $N_f$-fold degenerate flat band $f$, which is well-isolated from all other bands $n\neq f$ with $|E_{n,\bk}-E_f|\gg \Gamma$. We set the chemical potential to $\mu=E_f$
and obtain \cite{Mitscherling2022}
\begin{align}
    \label{eqn:sigmaFlat}
    \sigma^{ij,s}_\text{inter}= \frac{2}{\pi} \frac{e^2}{\hbar}\int\!\!\!\frac{d^d\bk}{(2\pi)^d}\,g^{ij}_{f}(\bk) \equiv \frac{2}{\pi} \frac{e^2}{\hbar} \overline g^{\,ij}_f\, ,
\end{align}
where we introduced the quantum metric $\overline g^{\,ij}_f$ of the flat band integrated over the Brillouin zone. The result in Eq.~\eqref{eqn:sigmaFlat} also holds for almost flat bands with $|E_{f,\bk}-\mu|\ll \Gamma$. We see that the dominant contribution to the longitudinal conductivity of the flat band is given by the quantum metric of the underlying Grassmannian, since the quasiparticle velocities $ \dki E_{f,\bk}=0$ ($\approx 0$) of an (almost) flat band is strongly suppressed \cite{Mitscherling2022}.

{\it Nontrivial degenerate flat bands.---} We construct a three-band toy model $H(\bk)$ with topologically nontrivial flat bands in two dimensions. Let us consider $\vec{n}_\bk=\vec{d}_\bk/|\vec{d}_\bk|$ with $\vec{d}_\bk=(\sin k_x,\cos k_y, 1-\cos k_x-\cos k_y)$. We use a spin-1 irreducible representation of SU(2), 
\begin{gather}
S_1\!=\!\frac{1}{\sqrt{2}}\!\!\left[
\begin{array}{ccc}
 \!0 & \!1 & \!0 \\
 \!1 & \!0 & \!1 \\
 \!0 & \!1 & \!0 \\
\end{array}
\right]\!\!,\, 
S_2\!=\!\frac{1}{\sqrt{2}}\!\!\left[
\begin{array}{ccc}
 \!0 & \!i & \!0 \\
 \!-i & \!0 & \!i \\
 \!0 & \!-i & \!0 \\
\end{array}
\right]\!\!,\, 
S_3\!=\!\left[\begin{array}{ccc}
 \!-1 & \!0 & \!0 \\
 \!0 & \!0 & \!0 \\
 \!0 & \!0 & \!1 \\
\end{array}\right]\!.
\label{eq: spin-1 irrep}
\end{gather}
in order to define the projectors
\begin{gather}
P_0(\bk)=1-h_\bk^2\,,\,P_\pm(\bk)=\frac{1}{2}\left[\pm h_\bk+h_\bk^2\right]\,,
\end{gather}
where $h_\bk=\vec{n}_\bk\cdot \vec{S}$ with $\vec{S}=(S_1,S_2,S_3)$. These projectors correspond to the three momentum-independent eigenvalues $0$ and $\pm 1$ of $h_\bk$ \cite{SupplMat}. We will use the three band energies $\varepsilon_n$ in 
\begin{align}
H(\bk)=\sum_{n=-,0,+}\varepsilon_{n}\,P_n(\bk) 
\label{eq: Bloch Hamiltonian spin-1 model}
\end{align}
to discuss the impact of degeneracy on the longitudinal conductivity of flat
bands. We have
$\sigma^{xx}_\text{intra}=0$
and calculate the longitudinal conductivity $\sigma^{xx}=\sigma^{xx,s}_{\text{inter}}$ via Eq.~\eqref{eqn:sigmaS} at zero temperature.

\begin{figure}[t!]
    \centering
    \includegraphics[width=0.45\textwidth]{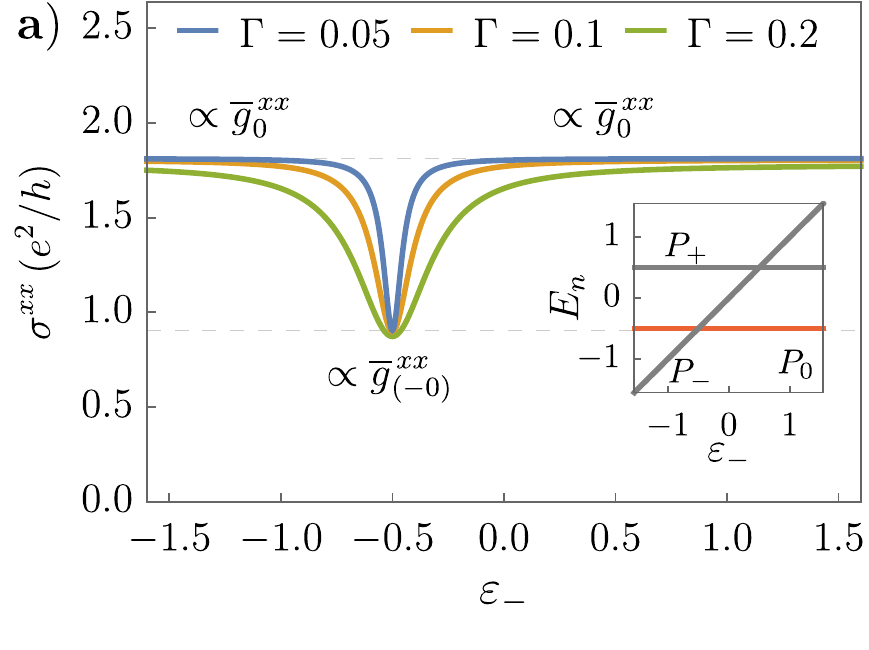}
    \includegraphics[width=0.45\textwidth]{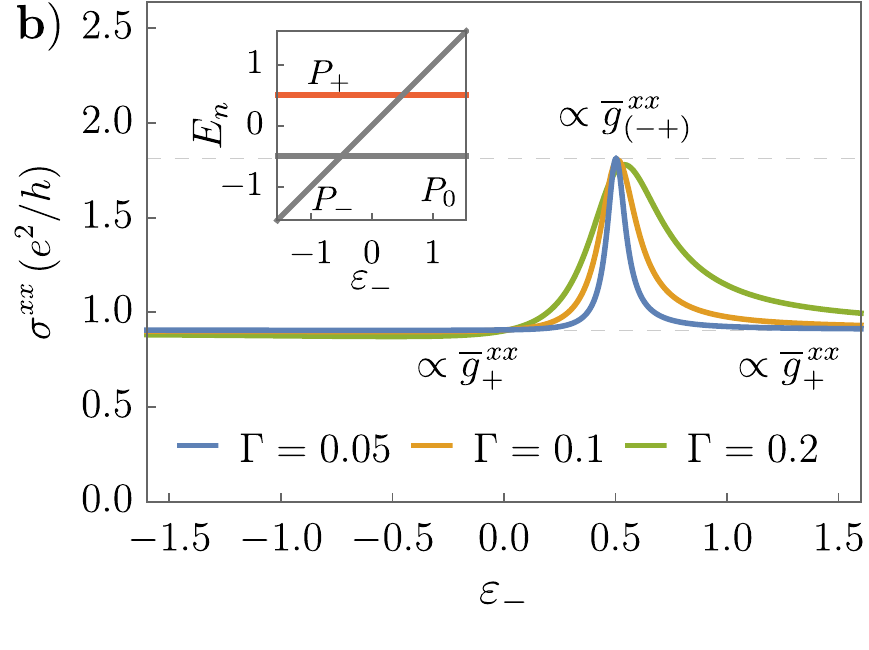}
    \caption{The longitudinal conductivity $\sigma^{xx}$ of a flat band is proportional to the corresponding integrated quantum metric $\overline g^{xx}_f$ (dashed lines). 
    When the bands become degenerate, we find a pronounced drop for $\mu=\varepsilon_0$ (a) and peak for $\mu = \varepsilon_+$ (b). 
    In the inset, we show the energy levels of the three bands.} 
    \label{fig:fig2}
\end{figure}

\begin{figure}[t!]
    \centering
    \includegraphics[width=0.45\textwidth]{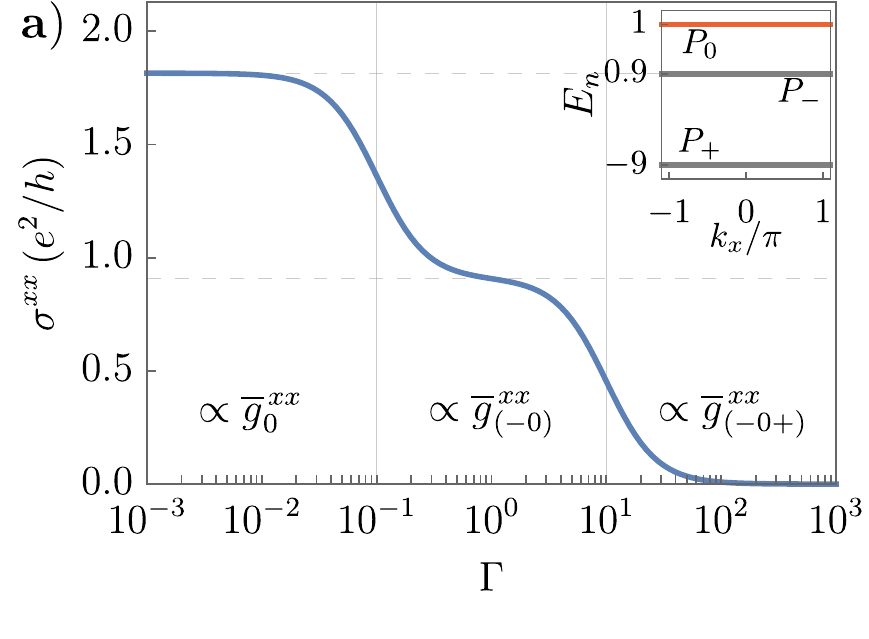}\\
    \includegraphics[width=0.45\textwidth]{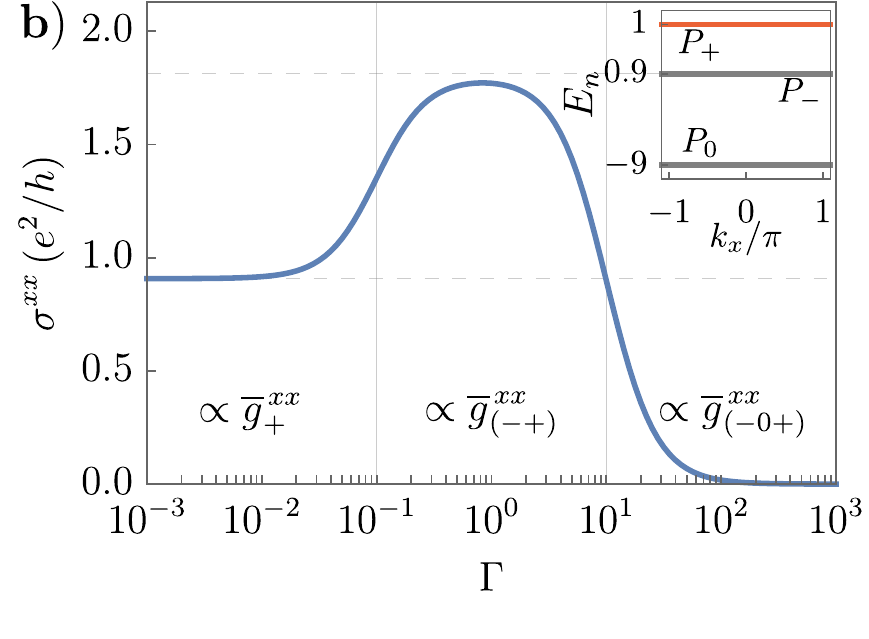}\\
    \includegraphics[width=0.45\textwidth]{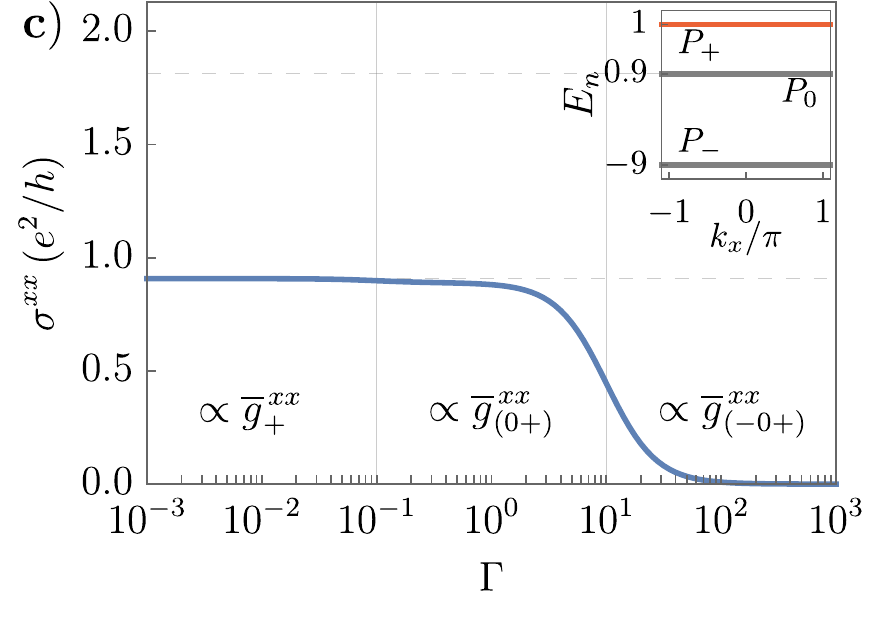}
    \caption{The longitudinal conductivity $\sigma^{xx}$ as a function of a phenomenological band broadening $\Gamma$ for different relative position of the three bands. 
    We understand the value of the observed plateaus (dashed lines) by the quantum metric of the involved one, two, or three bands (inset). The crossovers are given by the band gaps (vertical lines). 
    } \label{fig:fig3}
\end{figure}

In Fig.~\ref{fig:fig2}, we show  $\sigma^{xx}$ as a function of the energy level $\varepsilon_-$ for different $\Gamma$. 
We fix the chemical potential $\mu$ to the flat-band energies $\varepsilon_0=-0.5$ (a) and $\varepsilon_+=0.5$ (b). 
In Fig.~\ref{fig:fig2}(a), we find a drop of $\sigma^{xx}$ when $\varepsilon_-=\varepsilon_0$. In contrast, we find a peak when $\varepsilon_-=\varepsilon_+$ in Fig.~\ref{fig:fig2}(b). 
Via Eq.~\eqref{eqn:sigmaFlat}, we can relate this behavior to the different quantum metrics of nondegenerate and degenerate bands \cite{SupplMat}. If $|\varepsilon_{0/+}-\varepsilon_-|\gg \Gamma$, the flat band at energy $\varepsilon_{0/+}$ is isolated and nondegenerate. We have (a) $\sigma^{xx}=4\, \overline g^{\,xx}_{0}=4\,c$ and (b) $\sigma^{xx}=4\, \overline g^{\,xx}_{+}=2\,c$ in units $e^2/h$, where 
$c=\int\!\!\frac{d^2\bk}{(2\pi)^2}\partial_x \vec{n}_\bk\cdot \partial_x \vec{n}_\bk
\approx 0.454$.
If $|\varepsilon_{0/+}-\varepsilon_-|
\ll \Gamma$, the flat band is isolated and twofold degenerate. We have (a) $\sigma^{xx}=4\,\overline g^{\,xx}_{(-0)}=2\, c$ and (b) $\sigma^{xx}=4\,\overline g_{(+-)}^{\,xx}=4\,c$. 

In Fig.~\ref{fig:fig3}, we show $\sigma^{xx}$ as a function of the relevant energy scale $\Gamma$.
We fix $\mu=1$ to the highest band $1$. The band gap to the middle band $2$ and lowest band $3$ are $\Delta_{12}=0.1$ and $\Delta_{13}=10$, respectively.
Using Eq.~\eqref{eqn:sigmaFlat}, we can relate the obtained conductivity plateaus to the integrated quantum metric, i.e., $\sigma^{xx}=4\,\overline g_{1}^{\,xx}$ for $\Gamma\lesssim \Delta_{12}$ and $\sigma^{xx}=4\,\overline g_{(12)}^{\,xx}$ for $\Delta_{12}\lesssim \Gamma\lesssim \Delta_{13}$ in units $e^2/h$. Here, we have an effective two-fold degeneracy of bands 1 and 2 set by the scale $\Gamma>\Delta_{12}$. In agreement with Fig.~\ref{fig:fig2}, we recover the drop and rise of the conductivity in Figs.~\ref{fig:fig3}(a) and \ref{fig:fig3}(b), respectively. We have $\overline g_{+}^{\,xx}=\overline g_{(+0)}^{\,xx}=c/2$ \cite{SupplMat}, so that the conductivity does not change between the nondegenerate and degenerate band in Fig.~\ref{fig:fig3}(c). We discuss the crossover behaviors in the SM \cite{SupplMat}. 

{\it Superfluid stiffness.---} The step from Eq.~\eqref{eqn:sigmaS} to Eq.~\eqref{eqn:sigmaSdeg} and the application of identity \eqref{eqn:relation} can be analogously used for  
%for the quantities involving $g^{ij}_{n}$ or $\overline{g}^{\,ij}_{n}$ 
the superfluid stiffness tensor $D^{ij}_f$ of a degenerate flat band $f$~\cite{Peotta2015, Liang2017, Torma2021},  where we find
\begin{align}
D^{ij}_f=\frac{4 e^2 U\nu(1-\nu)}{\hbar^2}\int\!\!\!\frac{d^d\bf{k}}{\left(2\pi\right)^d}g^{ij}_f(\bf{k}),
\end{align}
with coupling strength $U$ and filling factor of the band $\nu$. Here, the relevant reference scale for the flat-band degeneracy is $U$ instead of the phenomenological relaxation rate $\Gamma$. Thus, we can directly apply Eqs.~\eqref{eqn:inequality} and \eqref{eqn:nonadditivity}. Inequalities of the form given in Eq.~\eqref{eqn:inequality} were used to derive lower bounds for the superfluid stiffness~\cite{Peotta2015}. The nonadditivity property of the quantum metric will manifest itself in the nonadditivity, under (effective) band collapse, of the superfluid stiffness $D^{ij}_f$. Our result is consistent with previous work where the importance of band degeneracy for the superfluid stiffness was noticed before \cite{Peotta2015}. It is crucial for twisted bilayer graphene \cite{Xie2020}.

We note that it has been shown recently that the so-called minimal quantum metric, the metric with minimal trace, should be considered when computing of the superfluid stiffness $D^{ij}_f$ \cite{Huhtinen2022}.

{\it Conclusions.---}
We have shown how the nonadditivity of the quantum metric upon collapse of a collection of bands manifests itself in physical observables (such as the dc electrical conductivity and the superfluid stiffness). We have given a physical interpretation for the term responsible for failure of additivity in terms of transition dipole matrix elements between two bands. We suggest that this distinguished property may be used to infer quantum metric effects. Furthermore, it provides a new purely quantum-geometrical mechanism for manipulating measurable quantities by changing the underlying degeneracy. It would be interesting to study the effect of nonadditivity of the quantum metric in disordered systems and in systems with interactions. Several direct measurements of the quantum metric have been reported recently \cite{Asteria2019, Tan2019, Gianfrate2020, Yu2020, Tan2021, Tian2021, Yu2021, Chen2022a}, which might serve as a good starting point for an experimental verification of this effect.

\begin{acknowledgments}
%We thank M.~M.~Hirschmann, T.~Holder, J.~E.~Moore, N.~Goldman and T.~Ozawa for carefully reading through this manuscript and for stimulating comments. 
We thank M.~M.~Hirschmann, T.~Ozawa and J.~E.~Moore for carefully reading through this manuscript and for stimulating comments. 
B.M. acknowledges very important discussions with T. Ozawa before the starting date of this work, where the peculiar differences between degenerate and nondegenerate band quantum metrics were pointed out. B.M. also acknowledges fruitful  discussions with N.~Goldman. J.M. thanks J.~Ahn, P.~M.~Bonetti, W.~Chen, T.~Holder, A.~Lau, A.~Leonhardt, W.~Metzner, and A.~Schnyder for stimulating discussions on the role of the quantum metric. 
We thank K.-E. Huhtinen for valuable discussions. J.M. acknowledges support by the German National Academy of Sciences Leopoldina through Grant No. LPDS 2022-06.
\end{acknowledgments}

B.M. and J.M. contributed equally to this work.

\bibliography{bib.bib}

\appendix

\begin{widetext}
\newpage
\section*{Supplemental Material}

In this Supplemental Material (SM), (i) we present some remarks on the differential geometry of band theory, (ii) we present the theory of dc electrical conductivity for degenerate bands in terms of the underlying quantum geometry, (iii) the discussion on gauge invariance of the relevant quantities in the context of degenerate bands, and (iv) provide two toy models on the collapse of bands including how the quantum geometry changes according to the degeneracy.

\section{Some remarks on the differential geometry of band theory}
We begin by presenting the derivation of Eq.~\eqref{eq: Berry curvature} in the main text. Let $\ket{u_{\bf{k}}}$ be a section of $\text{Im}(P)\subset\mathcal{E}$. By definition,
\begin{align}
P(\bf{k})\ket{u_{\bf{k}}}=\ket{u_{\bf{k}}}.
\end{align}
The Berry curvature is a $2$-form with values in the endomorphisms of $\text{Im}(P)$ and it acts on sections of $\text{Im}(P)$ by 
\begin{align}
\Omega\left(\ket{u_{\bf{k}}}\right)&= P(\bf{k})\nabla\wedge \left(P(\bf{k})\nabla \left(\ket{u_{\bf{k}}}\right)\right),
\end{align}
where $\nabla$ is the Berry connection on $\mathcal{E}$ which satisfies $\nabla\wedge \nabla=0$, i.e., it is \emph{flat}. We can then expand the right-hand side using the Leibniz rule,
\begin{align}
 P(\bf{k})dP(\bf{k})\wedge\nabla \left(\ket{u_{\bf{k}}}\right) + P(\bf{k})\nabla\wedge \nabla\left(\ket{u_{\bf{k}}}\right)&=   P(\bf{k})dP(\bf{k})\wedge\nabla \left(\ket{u_{\bf{k}}}\right) \nonumber \\
 &=P(\bf{k})dP(\bf{k})\wedge\nabla \left(P(\bf{k})\ket{u_{\bf{k}}}\right) \nonumber\\
 &=P(\bf{k})dP(\bf{k})\wedge dP(\bf{k}) \ket{u_{\bf{k}}}+\left(P(\bf{k})dP(\bf{k})P(\bf{k})\right)\wedge \nabla\ket{u_{\bf{k}}}, \nonumber\\
 &=\left(P(\bf{k})dP(\bf{k})\wedge dP(\bf{k})P(\bf{k}) \right)\ket{u_{\bf{k}}},
\end{align}
where we have used flatness of $\nabla$, $P(\bf{k})\ket{u_{\bf{k}}}=\ket{u_{\bf{k}}}$, $P(\bf{k})^2=P(\bf{k})$ and $P(\bf{k})dP(\bf{k})P(\bf{k})=0$ [which follows from differentiation of $P(\bf{k})^2=P(\bf{k})$, multiplication on the left and on the right by $P(\bf{k})$ and using $P(\bf{k})^2=P(\bf{k})$]. We then clearly see
\begin{align}
\Omega=PdP\wedge dPP.
\end{align}
Note that, unlike the connection, the curvature $\Omega$ is not a differential operator, but rather a vector bundle homomorphism with values in the $2$-forms.

Expressions that use $P(\bf{k})$ rather than (local) Bloch wave functions are useful because they express gauge invariance in a rather straightforward way. However, it is also convenient to have expression in terms of local Bloch wave functions. In particular, if one fixes a local orthonormal basis of sections of $\text{Im}(P)$ (i.e. a unitary frame field or a unitary gauge), $\ket{u_{m,\bf{k}}}$, $m=1,\dots,r$, where $r=\tr (P)$ is the rank of $\text{Im}(P)\to\BZ^d$, we can write
\begin{align}
P\nabla \ket{u_{m,\bf{k}}}=\sum_{n=1}^{r} \bra{u_{n,\bf{k}}} d\ket{u_{m,\bf{k}}}\ket{u_{n,\bf{k}}}
\end{align}
The $r\times r$ matrix $\theta=[\theta^{n}_{\; m}]_{1\leq n,m\leq r}=[\bra{u_{n,\bf{k}}}d\ket{u_{m,\bf{k}}}]_{1\leq n,m\leq r}$ is the local connection $1$-form, and we may write the previous equation as
\begin{align}
P\nabla \ket{u_{m,\bf{k}}}=\sum_{n=1}^{r}\theta^{n}_{\; m}\ket{u_{n,\bf{k}}}.
\end{align}
Observe that if $P(\bf{k})$ is an orthogonal projector describing a degenerate band of a Bloch Hamiltonian $H(\bf{k})$, then $\theta$ is the restriction of the $N\times N$ matrix $A$ in Eq.~\eqref{eqn:BerryConnection} of the main text to the $r\times r$ sub-matrix corresponding to the frame field $\ket{u_{m,\bf{k}}}$, $m=1,\dots,r$, which generates $\text{Im}(P(\bf{k}))\subset \mathcal{E}_{\bf{k}}$ and which corresponds to $r$ columns of the unitary matrix $U(\bf{k})$, locally diagonalizing $H(\bf{k})$.
Any section $\ket{u_{\bf{k}}}$ of the bundle $\text{Im}(P)$ may be written locally as
\begin{align}
\ket{u_{\bf{k}}}=\sum_{m=1}^{r}a^m(\bf{k})\ket{u_{m,\bf{k}}},
\end{align}
for locally defined smooth functions $a^1(\bf{k}),\dots, a^{r}(\bf{k})$. Using the Leibniz rule, it follows that the covariant derivative of $\ket{u_{\bf{k}}}$ is determined by
\begin{align}
P\nabla\left(\ket{u_{\bf{k}}}\right)=\sum_{m=1}^{r}\left(da^m(\bf{k}) +\sum_{n=1}^{r}\theta^{m}_{\; n}a^{n}(\bf{k})\right)\ket{u_{m,\bf{k}}}.
\end{align}
Applying the operator $P\nabla\wedge $, we find, using the Leibniz rule,
\begin{align}
\Omega\left(\ket{u_{m,\bf{k}}}\right)&=\left(P\nabla \wedge P\nabla\right)\left(\ket{u_{m,\bf{k}}}\right) \nonumber\\
&=\left(P\nabla\wedge\right)\left(\sum_{n=1}^{r}\theta^{n}_{\; m}\ket{u_{n,\bf{k}}}\right) \nonumber\\
&=\sum_{n=1}^{r}\left(d\theta^{n}_{\; m}+\sum_{p=1}^{r} \theta^{n}_{\; p}\wedge \theta^{p}_{\; m}\right)\ket{u_{n,\bf{k}}},
\end{align}
where we used anti-symmetry of the wedge product in the last line.
Writing $\Theta=[\Theta^{n}_{\; m}]_{1\leq m,n\leq r}$ for the matrix valued $2$-form whose components are
\begin{align}
\Theta^{n}_{\; m}=d\theta^{n}_{\; m}+\sum_{p=1}^{r} \theta^{n}_{\; p}\wedge \theta^{p}_{\; m},
\end{align}
we see that
\begin{align}
\Omega\left(\ket{u_{m,\bf{k}}}\right)=\sum_{n=1}^{r}\Theta^{n}_{\; m}\ket{u_{n,\bf{k}}}.
\end{align}
We conclude that the local forms of the covariant derivative and the associated curvature, in terms of the local unitary frame field $\ket{u_{1,\bf{k}}},\dots,\ket{u_{r,\bf{k}}}$ are, respectively, $d+\theta$ and $\Theta=d\theta+\theta\wedge \theta$. Finally, note that
\begin{align}
\Theta^{n}_{\; m} &= d\theta^{n}_{\; m} +\sum_{p=1}^{r} \theta^{n}_{\; p}\wedge \theta^{p}_{\; m} \nonumber \\
&=\bra{ du_{n,\bf{k}}}\wedge \ket{ du_{m,\bf{k}}} +\sum_{p=1}^{r}\bra{u_{n,\bf{k}}}d\ket{u_{p,\bf{k}}}\wedge \bra{u_{p,\bf{k}}}d\ket{u_{m,\bf{k}}}\nonumber\\
&=\bra{du_{n,\bf{k}}}\wedge \left(1-\sum_{p=1}^{r}\ket{u_{p,\bf{k}}}\bra{u_{p,\bf{k}}}\right)\ket{du_{m,\bf{k}}}\nonumber\\
&=\bra{du_{n,\bf{k}}}\wedge Q(\bf{k})\ket{du_{m,\bf{k}}}\nonumber\\
&=\sum_{i,j=1}^{d}\bra{\partial_i u_{n,\bf{k}}} Q(\bf{k})\ket{\partial_j u_{m,\bf{k}}}dk_i\wedge dk_j.
\end{align}
It then follows that
\begin{align}
F&=\tr\left(\Omega\right)=\tr\left(PdP\wedge dPP\right)=\tr\left(PdP\wedge dP\right)\nonumber\\
&=\sum_{n=1}^{r}\bra{u_{n,\bf{k}}}\Omega\ket{u_{n,\bf{k}}}=\sum_{n=1}^{r}\Theta^{n}_{\; n} \nonumber\\
&=\sum_{i,j=1}^{d}\sum_{m=1}^{r}\bra{\partial_i u_{n,\bf{k}}} Q(\bf{k})\ket{\partial_j u_{n,\bf{k}}}dk_i\wedge dk_j,
\end{align}
where we used the cyclic property of the trace and $P^2=P$ in the first line. This establishes Eq.~\eqref{eq: Abelian Berry curvature}.

\section{DC electrical conductivity of degenerate bands}
The dc conductivity tensor $\sigma^{ij}$ decomposes into \cite{Mitscherling2020}
\begin{align}
    \label{eqn:sigma}
    \sigma^{ij}=\sigma^{ij}_\text{intra}+\sigma^{ij,s}_\text{inter}+\sigma^{ij,a}_\text{inter}\, .
\end{align}
In the main text, we discuss only the quantum metric contribution $\sigma^{ij,s}_\text{inter}$. Here, we will present the complete discussion including the intraband contribution $\sigma^{ij}_\text{intra}$ and the Berry curvature contribution $\sigma^{ij,a}_\text{inter}$. For the non-interacting $N$-band Hamiltonian given in Eq.~\eqref{eqn:H}, we have \cite{Mitscherling2022}
\begin{align}
 &\sigma^{ij}_\text{intra}=\frac{e^2}{\hbar}\!\!\int\!\!\!\frac{d^d\bk}{(2\pi)^d} \sum_{n=1}^N w^{\text{intra}}_{n}(\bk)\,\, v^{i}_{n,\bk} v^{j}_{n,\bk} \, , \label{eqn:sigmaIntraAppendix}\\
 &\sigma^{ij,s}_\text{inter}=\frac{e^2}{\hbar}\!\!\int\!\!\!\frac{d^d\bk}{(2\pi)^d} \mathop{\sum_{n=1}^{N}\sum_{m=1}^{N}}_{n\neq m} w^{\text{inter},s}_{nm}(\bk)\,\, g^{ij}_{nm}(\bk) \, , \label{eqn:sigmaSAppendix}\\
 &\sigma^{ij,a}_\text{inter}=\frac{e^2}{\hbar}\!\!\int\!\!\!\frac{d^d\bk}{(2\pi)^d} \mathop{\sum_{n=1}^{N}\sum_{m=1}^{N}}_{n\neq m} w^{\text{inter},a}_{nm}(\bk)\,\, \Omega^{ij}_{nm}(\bk) \, , \label{eqn:sigmaAAppendix}
\end{align}
where $v^i_{n,\bk}=\partial_i E_{n,\bk}$ are the quasiparticle velocities, $g^{ij}_{nm}=\text{Re}\big[r^i_{nm}r^j_{mn}\big]$, and $\Omega^{ij}_{nm}=-2\text{Im}\big[r^i_{nm}r^j_{mn}\big]$, where $r^i_{nm}$ is the transition dipole matrix element defined in the main text. Each term is weighted by 
\begin{alignat}{2}
 &w^{\text{intra}}_{n}(\bk)&&=-\pi\!\!\int\!\!d\eps f'(\eps)\, \big[\mathcal{A}_n(\bf{k},\eps)\big]^2 \, ,\\
 &w^{\text{inter},s}_{nm}(\bk)&&=-\pi(E_{n,\bk}\!-\!E_{m,\bk})^2\!\!\int\!\!d\eps f'(\eps)\mathcal{A}_n(\bf{k},\eps)\mathcal{A}_m(\bf{k},\eps), \label{eqn:wS}\\
 &w^{\text{inter},a}_{nm}(\bk)&&=-\pi^2(E_{n,\bk}-E_{m,\bk})^2\!\!\int\!\!d\eps f(\eps) \Big(\big[\mathcal{A}_n(\bf{k},\eps)\big]^2\mathcal{A}_m(\bf{k},\eps)-\big[\mathcal{A}_m(\bf{k},\eps)\big]^2\mathcal{A}_n(\bf{k},\eps)\Big)\,, \label{eqn:wA}
\end{alignat}
where $\cA_n(\bk,\eps)$ is the quasiparticle spectral function of band $n$ defined in the main text. 

As in the main text, we consider $r$ bands, which are $N_i$-times degenerate. We have $\sum_{i=1}^r N_i=N$. We note that $w^{\text{intra}}_{ns}$, $w^{\text{inter},s}_{(ns)(ml)}$ and $w^{\text{inter},a}_{(ns)(ml)}$ only depend on the eigenenergies and are, thus, equal for all $s=1,...,N_n$ and $m=1,...,N_m$ belonging to the respective degenerate band $n$ and $m$. Using this, we split the summations in the conductivity contributions \eqref{eqn:sigmaIntraAppendix}, \eqref{eqn:sigmaSAppendix}, and \eqref{eqn:sigmaAAppendix}, and obtain
\begin{align}
 &\sigma^{ij}_\text{intra}=\frac{e^2}{\hbar}\!\!\int\!\!\!\frac{d^d\bk}{(2\pi)^d} \sum_{n=1}^r N^{}_n w^{\text{intra}}_{n}(\bk)\,\, v^{i}_{n,\bk} v^{j}_{n,\bk} \, , \label{eqn:sigmaIntraDeg}\\
 &\sigma^{ij,s}_\text{inter}=\frac{e^2}{\hbar}\!\!\int\!\!\!\frac{d^d\bk}{(2\pi)^d} \mathop{\sum_{n=1}^{r}\sum_{m=1}^{r}}_{n\neq m} w^{\text{inter},s}_{nm}(\bk)\,\, \widehat g^{\,ij}_{nm}(\bk) \, , \label{eqn:sigmaSDeg}\\
 &\sigma^{ij,a}_\text{inter}=\frac{e^2}{\hbar}\!\!\int\!\!\!\frac{d^d\bk}{(2\pi)^d} \mathop{\sum_{n=1}^{r}\sum_{m=1}^{r}}_{n\neq m} w^{\text{inter},a}_{nm}(\bk)\,\, \widehat{\Omega}^{\,ij}_{nm}(\bk) \, . \label{eqn:sigmaADeg}
\end{align}
We see that the intraband contribution $\sigma^{ij}_\text{intra}$ of a degenerate band is trivially enhanced by the factor $N_i>1$. In contrast, the interband contributions involve
\begin{align}
   \label{eqn:gnmAppendix} \widehat g^{\,ij}_{nm}(\bf{k}) &\equiv \sum_{r=1}^{N_n}\sum_{s=1}^{N_m} \Re\!\big[r^i_{(nr)(ms)}r^j_{(ms)(nr)}\big]=\sum_{r=1}^{N_n}\sum_{s=1}^{N_m} \Re\!\Big[i\langle u_{nr,\bk}|\partial_i u_{ms,\bk}\rangle i\langle u_{ms,\bk}|\partial_j u_{nr,\bk}\rangle\Big] \\
   \label{eqn:Onm} \widehat \Omega^{\,ij}_{nm} (\bf{k})&\equiv-2  \sum_{r=1}^{N_n}\sum_{s=1}^{N_m} \Im\!\big[r^i_{(nr)(ms)}r^j_{(ms)(nr)}\big]= -2\sum_{r=1}^{N_n}\sum_{s=1}^{N_m}\Im\! \Big[i\langle u_{nr,\bk}|\partial_i u_{ms,\bk}\rangle i\langle u_{ms,\bk}|\partial_j u_{nr,\bk}\rangle\Big] \, ,
\end{align}
which includes the remaining summations over the degenerate subspaces of the involved degenerate bands $n$ and $m$. In the next section, we will show that $\widehat g^{\, ij}_{nm}$ and $\widehat \Omega^{\, ij}_{nm}$ are $U(N_n)\times U(N_m)$-gauge invariant. We have seen in the main text that 
\begin{align}
    \sum_{\underset{m\neq n}{m=1}}^r \widehat g^{\,ij}_{nm}(\bk)=g^{ij}_{n}(\bf{k}) \, ,
    \label{eqn:relationG}
\end{align}
which relates the gauge-invariant transition rates $\widehat g^{\, ij}_{nm}$ to the quantum metric of the underlying Grassmannian manifold of band $n$. Analogously, we have 
\begin{align}
    \sum_{\underset{m\neq n}{m=1}}^r \widehat \Omega^{\,ij}_{nm}(\bk)=iF^{\,ij}_n(\bk) \, , 
\end{align}
so that we identify the Abelian Berry curvature of the underlying Grassmannian of band $n$.   

\section{Gauge invariance}

We show explicitly that $\widehat{g}^{\,ij}_{nm}$ and $\widehat{\Omega}^{\,ij}_{nm}$, which are defined in Eq.~\eqref{eqn:gnmAppendix} (or Eq.~\eqref{eqn:gnm} in the main text) and Eq.~\eqref{eqn:Onm}, respectively, are invariant under a $\text{U}(N_n)\times\text{U}(N_m)$-gauge transformation of the underlying degenerate bands $n$ and $m$. Let us assume a local unitary gauge transformations $U^p\in \text{U}(N_p)$, with $p=n,m$, where here and below, for the sake of simplicity, we omit the $\bf{k}$ dependence of the involved quanties, i.e., $U^p\equiv U^{p}(\bf{k})$, $\ket{u_{mr}}\equiv \ket{u_{mr,\bf{k}}}$ and $\ket{u_{ns}}\equiv \ket{u_{ns,\bf{k}}}$, with $r=1,\dots,N_{m}$ and $s=1,\dots,N_{n}$. Its components fulfill the identity
\begin{align}
    \sum_{\tilde s=1}^{N_p}(U^p_{\tilde s s})^*\,U^p_{\tilde s s'}=\delta^{}_{s s'}, \label{eqn:unitary}
\end{align}
where ${}^*$ denotes complex conjugation. We have
\begin{align}
    \ket{u_{ps}}=\sum^{N_p}_{\tilde s=1} U^p_{s\tilde s}\,\ket{u_{n\tilde s}} \, , \hspace{1cm}
    \bra{u_{ps}}=\sum^{N_p}_{\tilde s=1}(U^p_{s \tilde s})^*\,\bra{u_{n\tilde s}} \,,
\end{align}
with $p=n,m$.
Thus, the transition dipole moment transforms as
\begin{align}
    r^j_{(ns)(mr)}&=i\bra{u_{ns}}\partial_j\ket{u_{mr}}=\sum_{\tilde s=1}^{N_n}\sum_{\tilde r=1}^{N_m} i\, (U^n_{s\tilde s})^*\,\big(\partial_j U^m_{r\tilde r}\big)\,\,\langle u_{n\tilde s}|u_{m\tilde r}\rangle + \sum_{\tilde s=1}^{N_n}\sum_{\tilde r=1}^{N_m} (U^n_{s\tilde s})^*\, U^m_{r\tilde r}\,\, r^j_{(n\tilde s)(m\tilde r)} \label{eqn:rGauge2}
\end{align}
We see that the first term on the right hand side drops for $n\neq m$, since the subspaces are orthogonal to each other. Finally, we calculate
\begin{align}
    \sum_{s=1}^{N_n}\sum_{r=1}^{N_m}r^i_{(ns)(mr)}r^j_{(mr)(ns)} &=\sum_{s,\tilde s, \tilde s'=1}^{N_n}\sum_{r,\tilde r,\tilde r'=1}^{N_m}(U^n_{s\tilde s})^*\, U^m_{r\tilde r}\,\, (U^m_{r\tilde r'})^*\, U^n_{s\tilde s'} r^i_{(n\tilde s)(m\tilde r)}\,\,r^j_{(m\tilde r')(n\tilde s')}\\&=\sum_{\tilde s, \tilde s'=1}^{N_n}\sum_{\tilde r,\tilde r'=1}^{N_m} \delta_{\tilde s,\tilde s'}\delta_{\tilde r,\tilde r'} r^i_{(n\tilde s)(m\tilde r)}\,\,r^j_{(m\tilde r')(n\tilde s')}  \\&=\sum_{\tilde s=1}^{N_n}\sum_{\tilde r=1}^{N_m} r^i_{(n\tilde s)(m\tilde r)}\,\,r^j_{(m\tilde r)(n\tilde s)} \, ,
\end{align}
where we used \eqref{eqn:unitary} in the second step. After taking the real and imaginary part, we see that $\widehat g^{\,ij}_{nm,\bk}$ and $\widehat \Omega^{\,ij}_{nm,\bk}$ as defined in \eqref{eqn:gnmAppendix} and \eqref{eqn:Onm} are gauge invariant under $\text{U}(N_n)\times\text{U}(N_m)$ of the coupled degenerate subspaces.

An alternative proof of this result can be given as follows. One introduces orthogonal projectors $P_m=\sum_{r=1}^{N_{m}}\ket{u_{mr}}\bra{u_{mr}}$ and $P_n=\sum_{s=1}^{N_{n}}\ket{u_{ns}}\bra{u_{ns}}$. The quantity
\begin{align}
\tr\left( P_n \frac{\partial P_m}{\partial k_i} P_m\frac{\partial P_m}{\partial k_j} P_n\right)
\end{align}
is gauge invariant (it is written solely in terms of the orthogonal projectors which are gauge invariant themselves) and due to $P_n$ and $P_m$ defining at each $\bf{k}$ mutually orthogonal subspaces, it follows that
\begin{align}
\tr\left( P_n \frac{\partial P_m}{\partial k_i} P_m\frac{\partial P_m}{\partial k_j} P_n\right)&=\tr\left(\sum_{s,s'=1}^{N_n}\sum_{r=1}^{N_m} \ket{u_{ns}}\bra{u_{ns}}\frac{\partial}{\partial k_i}\ket{u_{mr}}\frac{\partial}{\partial k_j}\left(\bra{u_{mr}}\right)\ket{u_{ns'}}\bra{u_{ns'}}\right) \nonumber \\
&=\sum_{s=1}^{N_n}\sum_{r=1}^{N_m}\bra{u_{ns}}\frac{\partial}{\partial k_i}\ket{u_{mr}}\frac{\partial}{\partial k_j}\left(\bra{u_{mr}}\right)\ket{u_{ns}}\nonumber\\
&=\sum_{s=1}^{N_n}\sum_{r=1}^{N_m}\bra{u_{ns}}i\frac{\partial}{\partial k_i}\ket{u_{mr}}\bra{u_{mr}}i\frac{\partial}{\partial k_j}\ket{u_{ns}}=\sum_{s=1}^{N_n}\sum_{r=1}^{N_m}r^i_{(ns)(mr)}r^j_{(mr)(ns)}.
\end{align}
As in the last part of the previous proof, taking real and imaginary parts produces the desired result.

\section{Toy models for collapsing bands}
\subsection{(Trivial) Example of two band collapse and different metrics}
Suppose we have the two-band Hamiltonian
\begin{align}
H(\bf{k})=\varepsilon_{1,\bk}\, p(\bk) +\varepsilon_{2,\bk}\, q(\bk),
\end{align}
where
\begin{align}
p(\bk)=\frac{1}{2}\left(1+\vec{n}_{\bk}\cdot \vec{\sigma}\right) \text{ and } q(\bk)=1-p(\bk),
\end{align}
and $\vec{n}:\BZ^d\to S^2; \bk\mapsto \vec{n}_{\bk}$ describes a vector in the Bloch sphere. Observe that $p(\bk)+q(\bk)=1$. This means that when the two bands collapse, i.e., when $\varepsilon_{1,\bk}=\varepsilon_{2,\bk}$, we get a trivial projector and the quantum metric is zero. However, each band has a quantum metric
\begin{align}
g=\frac{1}{2}\tr\left( dp dp\right)=\frac{1}{2}\tr\left( dq dq\right)=\frac{1}{4}d\vec{n}\cdot d\vec{n},
\end{align}
where we used $dq=-dp$.
The reason why the quantum metric associated with the collapsed rank $2$ projector is zero can be traced back to the mixed term contribution $\tr\left(dp dq\right)=-\tr\left(dpdp\right)=2g$
\begin{align}
\frac{1}{2}\tr\left[ d\left( p+q\right) d\left(p +q\right)\right]=\frac{1}{2}\tr\left(dpdp\right) +\frac{1}{2}\tr\left(dqdq\right) +\tr\left(dp dq\right)=2g-2g=0.
\end{align}
This construction illustrates how the quantum metric of a degenerate band constructed from two bands which collapse into one is not, in general, simply the sum of the quantum metrics associated to each band. 

More generally, we can consider a band projector,
\begin{align}
P(\bf{k})=P_1(\bf{k})+P_2(\bf{k}),
\end{align}
with $P_1$ and $P_2$ having ranks $r_1$ and $r_2$, respectively, with $r=r_1+r_2$ and $P_iP_j=\delta_{ij}P_j$, $i,j=1,2$. We consider
\begin{align}
H(\bf{k})=\varepsilon_{1,\bk}\, P_1(\bk) +\varepsilon_{2,\bk}\, P_2(\bk),
\end{align}
and collapse the two bands described by $P_1$ and $P_2$ by setting $\varepsilon_{1,\bf{k}}=\varepsilon_{2,\bf{k}}$. Provided $P_1+P_2$ is a constant projector, i.e., if $\text{Im}(P)$ is a trivial bundle, we have $dP_1=-dP_2$. Then the quantum metric of the composite band vanishes identically, since
\begin{align}
g=g_1+g_2+\tr(dP_1dP_2)\,,
\end{align}
where $g_i=(1/2)\tr(dP_idP_i)$, $i=1,2$ and 
the mixed term satisfies
\begin{align}
\tr(dP_1dP_2)=-\tr(dP_1dP_1)=-\tr(dP_2dP_2),
\end{align}
so that it cancels the contributions from each individual band since $g_1=(1/2)\tr(dP_1dP_1)=(1/2)\tr(dP_2dP_2)=g_2$.

\subsection{Collapsing bands within a $3$-band system}
Take the spin-$1$ (unitary) irreducible representation of $\text{SU}(2)$, denoted below by $\rho$, and consider the Hamiltonian
\begin{align}
H(\vec{n})=\vec{n}\cdot \vec{S},
\label{eq: spin 1 model}
\end{align}
where $\vec{S}=(S_1,S_2,S_3)$ are the generators of the Lie algebra in this representation and we choose $S_3$ to be diagonal. In particular, we can take
\begin{align}
S_1=\left[
\begin{array}{ccc}
 0 & \frac{1}{\sqrt{2}} & 0 \\
 \frac{1}{\sqrt{2}} & 0 & \frac{1}{\sqrt{2}} \\
 0 & \frac{1}{\sqrt{2}} & 0 \\
\end{array}
\right], \ S_2=\left[
\begin{array}{ccc}
 0 & \frac{i}{\sqrt{2}} & 0 \\
 -\frac{i}{\sqrt{2}} & 0 & \frac{i}{\sqrt{2}} \\
 0 & -\frac{i}{\sqrt{2}} & 0 \\
\end{array}
\right] \text{ and } S_3=\left[\begin{array}{ccc}
 -1 & 0 & 0 \\
 0 & 0 & 0 \\
 0 & 0 & 1 \\
\end{array}\right].
\label{eq: spin-1 irrep appendix}
\end{align}
There exists an element of $U$ of $\text{SU}(2)$ such that $\rho(U)$ rotates $H(\vec{n})$ into diagonal form
\begin{align}
H(\vec{n})=\rho(U)S_3\rho(U)^{\dagger}.
\end{align}
In fact, $U$ is only defined up to multiplication on the right by elements of the diagonal $\text{U}(1)$ subgroup and its class, under the quotient $\text{SU}(2)/\text{U}(1)\cong S^2$, determines $\vec{n}\in S^2$ uniquely. Since 
\begin{align}
S_3=\text{diag}(-1,0,1),
\end{align}
we find three eigenvector bundles over $S^2$, denoted as $E_{-}\to S^2$, $E_{0}\to S^2$, and $E_{+}\to S^2$, whose fibers over $\vec{n}$ are the corresponding eigenspaces of $H(\vec{n})$. 

\subsubsection{Quantum metric}

We wish to calculate the quantum metric of each of the bands. It is clear that for any $R\in\text{SO}(3)$ each of the projectors satisfies
\begin{align}
P(R\vec{n})=\rho(U)P(\vec{n})\rho(U)^{-1},
\label{eq: rotation covariance}
\end{align}
where $U$ is any element of $\text{SU}(2)$ with $U\left(\vec{n}\cdot \vec{\sigma}\right)U^{-1}=\left(R\vec{n}\right)\cdot \vec{\sigma}$. There are only two such elements $U$ and $-U$. The map $\text{SU}(2)\to\text{SO}(3)$ described above is just the usual double cover map. Since the Fubini-Study metric is invariant under unitary transformations, it follows that the quantum metric of these three bands has to be $\text{SO}(3)$-invariant. Therefore, up to a scale factor, they have to be of the form $d\vec{n}\cdot d\vec{n}$. To determine the precise scale factors, it is enough to compute the metric at $\vec{n}_0=(0,0,1)$. At that point, the fibers of $E_-$, $E_0$, and $E_{+}$ are spanned, respectively, by
\begin{align}
\ket{\psi_{-}(\vec{n}_0)}=\left[\begin{array}{c}
1\\
0\\
0
\end{array}\right],\ \ket{\psi_{0}(\vec{n}_0)}=\left[\begin{array}{c}
0\\
1\\
0
\end{array}\right]
 \text{ and } \ket{\psi_{+}(\vec{n}_0)}=\left[\begin{array}{c}
0\\
0\\
1
\end{array}\right].
\end{align}
We also have the associated orthogonal projectors $P_m(\vec{n}_0)=\ket{\psi_m(\vec{n}_0)}\bra{\psi_m(\vec{n}_0)}$, $m=-,0,+$. We will now consider the quantum metric $g_{i}$ associated to each of the $P_i$ and also the quantum metric $g_{(ij)}$ associated to $P_{ij}=P_i+P_j$, $i\neq j$ with $i,j\in \{-,0,+\}$. The one corresponding to the three bands is trivially zero because the corresponding projector is just the identity matrix. Due to the previous argument based on rotation invariance, see Eq.~\eqref{eq: rotation covariance}, it is enough to compute the scale factor at $\vec{n}_0$. To do so, we first recall the identity
\begin{align}
\bra{\psi_i(\vec{n})}d\ket{\psi_j(\vec{n})}=\frac{\bra{\psi_i(\vec{n})}dH\ket{\psi_j(\vec{n})}}{\varepsilon_j-\varepsilon_i}=\frac{\bra{\psi_i(\vec{n})}d\vec{n}\cdot \vec{S}\ket{\psi_j(\vec{n})}}{\varepsilon_j-\varepsilon_i},
\end{align}
where $\varepsilon_j$, $j=-,0,+$, denote the eigenvalues of $H$. Then, we also recall that, using $PdP=dPQ$ for any projector and $Q=I-P$ its orthogonal complement, we can write
\begin{align}
g_i(\vec{n})=\tr\left(P_idP_idP_i\right)=\tr\left(dP_iQ_idP_i\right)=\sum_{j\neq i}\frac{|\bra{\psi_i(\vec{n})}d\vec{n}\cdot \vec{S}\ket{\psi_j(\vec{n})}|^2}{\left(\varepsilon_j-\varepsilon_i\right)^2}
\end{align}
and
\begin{align}
g_{(ij)}(\vec{n})=\tr\left(P_{ij}dP_{ij}dP_{ij}\right)=\tr\left(dP_{ij}Q_{ij}dP_{ij}\right)=\sum_{k\notin \{i,j\}}\sum_{l\in \{i,j\}}\frac{|\bra{\psi_l(\vec{n})}d\vec{n}\cdot \vec{S}\ket{\psi_k(\vec{n})}|^2}{\left(\varepsilon_l-\varepsilon_k\right)^2}.
\end{align}
At this point, we note that any tangent vector to $S^2$ at $\vec{n}_0$ assumes the form $\vec{u}=(u_1,u_2,0)$, and hence, at $\vec{n}_0$,
\begin{align}
d\vec{n}\cdot \vec{S}=dn_1\,S_1+dn_2\, S_2 =\frac{1}{2}\left(dn\, S_{-}+d\overline{n}\, S_{+}\right),
\end{align}
where we introduced $dn=dn_1+idn_2$, and the ladder operators $S_{\pm}=S_1\pm iS_2$. We can then explicitly compute, at $\vec{n}=\vec{n}_0$,
\begin{align}
\frac{\bra{\psi_i(\vec{n})}d\vec{n}\cdot \vec{S}\ket{\psi_j(\vec{n})}}{\varepsilon_j-\varepsilon_i}=\frac{1}{2}\frac{\bra{\psi_i(\vec{n})}\left(dn\, S_{-}+d\overline{n}\, S_{+}\right)\ket{\psi_j(\vec{n})}}{\varepsilon_j-\varepsilon_i}=\frac{1}{2}\left(\delta_{j,i+1}\, dn \bra{\psi_i}S_{-}\ket{\psi_j} -\delta_{j,i-1}\,d\overline{n}\bra{\psi_i}S_{+}\ket{\psi_j}\right),
\end{align}
where one has to be careful since the states with $j=i+1$ and $j=i-1$ may not exist, in which case we are instructed to give the value zero. From this we can see that
\begin{align}
\frac{|\bra{\psi_i(\vec{n})}d\vec{n}\cdot \vec{S}\ket{\psi_j(\vec{n})}|^2}{\left(\varepsilon_j-\varepsilon_i\right)^2}&=\frac{1}{4}\left(\delta_{j,i+1} \,dn \bra{\psi_i}S_{-}\ket{\psi_j} -\delta_{j,i-1}\,d\overline{n}\bra{\psi_i}S_{+}\ket{\psi_j}\right)\left(-\delta_{i,j+1} \,dn \bra{\psi_j}S_{-}\ket{\psi_i} +\delta_{i,j-1}\,d\overline{n}\bra{\psi_j}S_{+}\ket{\psi_i}\right)\nonumber\\
&=\frac{|dn|^2}{4}\left(\delta_{j,i+1}|\bra{\psi_i}S_{-}\ket{\psi_{j}}|^2 +\delta_{j,i-1}|\bra{\psi_i}S_{+}\ket{\psi_{j}}|^2\right)\nonumber\\
&=\frac{d\vec{n}\cdot d\vec{n}}{4}\left(\delta_{j,i+1}|\bra{\psi_i}S_{-}\ket{\psi_{j}}|^2 +\delta_{j,i-1}|\bra{\psi_i}S_{+}\ket{\psi_{j}}|^2\right).
\end{align}
Hence,
\begin{align}
g_{i}=\frac{d\vec{n}\cdot d\vec{n}}{4}\sum_{j\neq i}\left(\delta_{j,i+1}|\bra{\psi_i}S_{-}\ket{\psi_{j}}|^2 +\delta_{j,i-1}|\bra{\psi_i}S_{+}\ket{\psi_{j}}|^2\right),
\end{align}
and, similarly,
\begin{align}
g_{(ij)}=\frac{d\vec{n}\cdot d\vec{n}}{4}\sum_{k\notin \{i,j\}}\sum_{l\in \{i,j\}}\left(\delta_{l,k+1}|\bra{\psi_l}S_{-}\ket{\psi_{k}}|^2 +\delta_{l,k-1}|\bra{\psi_l}S_{+}\ket{\psi_{k}}|^2\right).
\end{align}
Evaluating the sums explicitly, we get the final result, valid at any $\vec{n}\in S^2$,
\begin{align}
g_{-}=\frac{d\vec{n}\cdot d\vec{n}}{2},\ g_{0}=d\vec{n}\cdot d\vec{n}, \text{ and } g_{+}=\frac{d\vec{n}\cdot d\vec{n}}{2},
\label{eq: nondegenerate qmetrics appendix}
\end{align}
and for the rank $2$ projectors
\begin{align}
g_{(-0)}=\frac{d\vec{n}\cdot d\vec{n}}{2},\ g_{(+-)}=d\vec{n}\cdot d\vec{n}, \text{ and } g_{(+0)}=\frac{d\vec{n}\cdot d\vec{n}}{2}.
\label{eq: degenerate qmetrics appendix}
\end{align}

\subsubsection{Abelian Berry curvature}

For the Abelian Berry curvatures given by $\tr\left( PdP\wedge dP\right)$, where $P$ is any of the projectors considered, the calculation is similar. We compute at $\vec{n}_0$
\begin{align}
\frac{\bra{\psi_i(\vec{n})}d\vec{n}\cdot \vec{S}\ket{\psi_j(\vec{n})}\wedge \bra{\psi_j(\vec{n})}d\vec{n}\cdot \vec{S}\ket{\psi_i(\vec{n})}}{\left(\varepsilon_j-\varepsilon_i\right)^2}&=\frac{dn \wedge  d\overline{n}}{4}\left(\delta_{j,i+1}|\bra{\psi_i}S_{-}\ket{\psi_{j}}|^2 -\delta_{j,i-1}|\bra{\psi_i}S_{+}\ket{\psi_{j}}|^2\right).
\end{align}
We then use $dn\wedge d\overline{n}=-2i dn_1\wedge dn_2=-2i \vec{n}\cdot \left(d\vec{n}\times d\vec{n}\right)$, which holds at $\vec{n}=\vec{n}_0$. Thus, the Abelian Berry curvatures at $\vec{n}_0$
\begin{align}
F_{i}=-\frac{i}{2} \vec{n}\cdot \left(d\vec{n}\times d\vec{n}\right)\sum_{j\neq i}\left(\delta_{j,i+1}|\bra{\psi_i}S_{-}\ket{\psi_{j}}|^2 - \delta_{j,i-1}|\bra{\psi_i}S_{+}\ket{\psi_{j}}|^2\right),
\end{align}
and, similarly,
\begin{align}
F_{(ij)}=-\frac{i}{2} \vec{n}\cdot \left(d\vec{n}\times d\vec{n}\right)\sum_{k\notin \{i,j\}}\sum_{l\in \{i,j\}}\left(\delta_{l,k+1}|\bra{\psi_l}S_{-}\ket{\psi_{k}}|^2 -\delta_{l,k-1}|\bra{\psi_l}S_{+}\ket{\psi_{k}}|^2\right).
\end{align}
Note that the Abelian curvatures associated with the orthogonal projector $P_{ij}=P_i+P_j$ are simply the sums of those from $P_i$ and $P_j$---this is different from the case of the quantum metric, where there is a mixed term that does not cancel (here it does due to skew-symmetry of the wedge product). Evaluating the sums, this time, one gets
\begin{align}
F_{-}=-i \vec{n}\cdot \left(d\vec{n}\times d\vec{n}\right),\ F_{0}=0\text{ and } F_{+}=i \vec{n}\cdot \left(d\vec{n}\times d\vec{n}\right),
\label{eq: nondegenerate curv}
\end{align}
and for the rank $2$ projectors
\begin{align}
F_{(-0)}=-i\vec{n}\cdot \left(d\vec{n}\times d\vec{n}\right),\ F_{(+-)}=0 \text{ and } F_{(+0)}=i\vec{n}\cdot \left(d\vec{n}\times d\vec{n}\right).
\label{eq: degenerate curv}
\end{align}
By rotation invariance, this is the correct form for general $\vec{n}$. Using $\int_{S^2}\vec{n}\cdot\left(d\vec{n}\times d\vec{n}\right)=4\pi$, one immediately gets that the first Chern numbers of the line bundles $E_{-},E_{0},E_{+}$ are, respectively,
\begin{align}
\int_{S^2}\frac{iF_{-}}{2\pi}=2,\ \int_{S^2}\frac{iF_{0}}{2\pi}=0 \text{ and} \int_{S^2}\frac{iF_{+}}{2\pi}=-2.
\end{align}

\subsubsection{Construction of the Hamiltonian used in the main text}

With the above construction in mind, together with a map $\vec{n}:\BZ^2\to S^2$, we can build a Bloch Hamiltonian of the form
\begin{align}
H(\bk)=\sum_{m=-,0,+}\varepsilon_{n,\bk}\,P_n(\bk)=\varepsilon_{-,\bk}\,P_{-}(\bk)+\varepsilon_{0,\bk}\,P_{0}(\bk)+\varepsilon_{+,\bk}\,P_{+}(\bk),
\label{eq: Bloch Hamiltonian spin-1 model appendix}
\end{align}
where the $P_{m}(\bf{k}):=P_{m}(\vec{n}(\bk))$ are simply pullbacks, i.e. composition with the map $\bk\mapsto \vec{n}(\bk)$, of the $P_m$ defined for the Hamiltonian of Eq.~\eqref{eq: spin 1 model} and the $\varepsilon_{m,\bk}$ are energy parameter functions, which we can tune. In particular, we can take the energy bands to be flat, and collapse bands by manipulating the energy values. We determine the quantum metrics by pullback under the map $\bk\mapsto \vec{n}(\bk)$ of Eq.~\eqref{eq: nondegenerate qmetrics appendix} and Eq.~\eqref{eq: degenerate qmetrics appendix}. Collapsing the three bands yields a vanishing quantum metric.

Taking into account that the Berry curvatures are obtained via pullback by the map $\vec{n}:\bk\mapsto \vec{n}(\bk)$, the first Chern numbers of each of the bands are obtained by computing the degree (also known as the winding number) of the map $\vec{n}$, which is given by
\begin{align}
\deg(\vec{n})=\int_{\BZ^2}\frac{d^2\bk}{4\pi} \vec{n}\cdot\left(\frac{\partial \vec{n}}{\partial k_1}\times\frac{\partial \vec{n}}{\partial k_2}\right) \in\mathbb{Z}.
\end{align}
In particular we get $2\deg(\vec{n}),0,-2\deg(\vec{n})$ for the bands labelled by $-,0,+$, respectively, and, similarly, $2\deg(\vec{n}),0,-2\deg(\vec{n})$ for the degenerate bands labelled by $-0,+-,+0$, respectively. The sum of all bands has zero 1st Chern number because the resulting vector bundle is trivial.

\subsubsection{Expressions for the orthogonal projectors}
We have used the following simple expressions for the orthogonal projectors associated to the Hamiltonian in Eq.~\eqref{eq: spin 1 model}
\begin{align}
P_{-}(\vec{n})=\frac{1}{2}\left[-\left(\vec{n}\cdot\vec{S}\right)+\left(\vec{n}\cdot\vec{S}\right)^2\right],\ P_0(\vec{n})=1-\left(\vec{n}\cdot \vec{S}\right)^2 \text{and } P_{+}(\vec{n})=\frac{1}{2}\left[\left(\vec{n}\cdot\vec{S}\right)+\left(\vec{n}\cdot\vec{S}\right)^2\right].
\end{align}
The derivation of this formula can be done as follows. From the Green's function
\begin{align}
G(z;\vec{n})=\left(z-H(\vec{n})\right)^{-1}=\sum_{i}\frac{1}{z-\varepsilon_i}P_i(\vec{n}),
\end{align}
it follows that
\begin{align}
P_i(\vec{n})=\int_{\gamma_{i}}\frac{dz}{2\pi i}G(z;\vec{n}),
\end{align}
where $\gamma_i$ is a small contour enclosing only the isolated eigenvalue $\varepsilon_i$ counterclockwise. We use the Cayley-Hamilton theorem, which states that evaluating the characteristic polynomial on the matrix yields zero, to show
\begin{align}
\left(\vec{n}\cdot \vec{S}\right)^{3}=\vec{n}\cdot \vec{S}.
\end{align}
We write
\begin{align}
G(z;\vec{n})&=(z-H(\vec{n}))^{-1}=\frac{1}{z}\sum_{k=0}^{\infty}\left(\frac{H(\vec{n})}{z}\right)^k=\frac{1}{z}\left[1+ \left(z^{-1}+z^{-3}+\dots\right)\left(\vec{n}\cdot\vec{S}\right) +(z^{-2}+z^{-4}+\dots)\left(\vec{n}\cdot\vec{S}\right)^2\right]\nonumber\\
&=\frac{1}{z} +\frac{1}{z^2-1}\left(\vec{n}\cdot\vec{S}\right)+\frac{1}{z}\frac{1}{z^2-1}\left(\vec{n}\cdot\vec{S}\right)^2.
\end{align}
Evaluating the residues at the eigenvalues $-1,0,+1$ yields the desired result.
\subsubsection{Vanishing of $\bra{\psi_{+}(\vec{n})}d\ket{\psi_{-}(\vec{n})}$}
We give a proof that $\bra{\psi_{+}(\vec{n})}d\ket{\psi_{-}(\vec{n})}=0$ for any choice of local eigenvectors $\ket{\psi_{+}(\vec{n})}$ and $\ket{\psi_{-}(\vec{n})}$, with eigenvalues $+1$ and $-1$, respectively, of the Hamiltonian in Eq.~\eqref{eq: spin 1 model}. The proof is a consequence of the form of the matrices in Eq.~\eqref{eq: spin-1 irrep}, which form an irreducible representation of $\text{SU}(2)$. As stated above, if $U$ is a matrix diagonalizing $\vec{n}\cdot \vec{\sigma}$, then $\widetilde{U}=\rho(U)$ diagonalizes $H(\vec{n})$. As a consequence, $H(\vec{n})$ is diagonalized by elements in the representation of $\text{SU}(2)$ specified by the matrices in Eq.~\eqref{eq: spin-1 irrep}. By finding an $\text{SU}(2)$ matrix $U(\vec{n})$ diagonalizing $\vec{n}\cdot \vec{\sigma}$ locally on $S^2$, we then have a local unitary matrix $\widetilde{U}=\rho(U)$, whose columns are choices for the local eigenstates $\ket{\psi_{-}(\vec{n})},\ket{\psi_{0}(\vec{n})},\ket{\psi_{+}(\vec{n})}$ with eigenvalues $-1,0,+1$, respectively [note that this can only be done locally because the line bundles involved are non-trivial]. It follows that the matrix
\begin{align}
\widetilde{U}^{-1}d\widetilde{U}=\left[\bra{\psi_{i}(\vec{n})}d\ket{\psi_{j}(\vec{n})}\right]_{i,j\in\{-,0,+\}}
\end{align}
is a linear combination of the generators of the considered repesentation of $\text{SU}(2)$, i.e., a linear combination of the matrices in Eq.~\eqref{eq: spin-1 irrep}. Looking at Eq.~\eqref{eq: spin-1 irrep}, we see that the matrix elements $\left(S_{i}\right)_{+-}=0$, for all $i\in\{1,2,3\}$, and, since the choice of $\widetilde{U}=\rho(U)$ was arbitrary, the proof is complete. This has as a consequence that, for any Bloch Hamiltonian of the form of Eq.~\eqref{eq: Bloch Hamiltonian spin-1 model appendix}, the transition matrix dipole of $r^{j}_{+-}(\bk)=i\bra{\psi_{+}(\vec{n}(\bk))}\frac{\partial}{\partial k_j}\ket{\psi_{-}(\vec{n}(\bk))}$, vanishes for all $j=1,\dots,d$. Note that, for the same reason, the $r^{j}_{-+}(\bk)$ vanishes, too.

\subsubsection{Explicit form of the conductivity}

In the following, we explicitly calculate each term of the quantum metric contribution $\sigma^{xx,s}_\text{inter}$ in Eq.~\eqref{eqn:sigmaSdeg} of the main text, in order to identify the particular behavior, which we discussed in Fig.~\ref{fig:fig2} and \ref{fig:fig3}. We define the quantity
\begin{align}
    c^{ij}\equiv\int\!\!\frac{d^2\bk}{(2\pi)^2}\frac{\partial \vec{n}_\bk}{\partial k_i}\cdot \frac{\partial \vec{n}_\bk}{\partial k_j},\ i,j\in\{x,y\},
\end{align}
and note that we have $c^{xx}=c^{yy}\equiv c$ and $c^{xy}=c^{yx}=0$ for the model under consideration. In particular, we have
\begin{align}
   c=\int\!\!\frac{d^2\bk}{(2\pi)^2}\partial_x \vec{n}_\bk\cdot \partial_x \vec{n}_\bk=\frac{1}{4\pi}\big[\pi-E(-8)+7K(-8)\big]
\approx 0.454\,. 
\label{eqn:valueC}
\end{align}
Here, $K(m)=\int_0^{\pi/2} \big(1-m\sin^2\theta\big)^{-1/2}d\theta$ and $E(m)=\int_0^{\pi/2} \big(1-m\sin^2\theta\big)^{1/2}d\theta$ are the complete elliptic integrals of the first and second kind, respectively. We define the transition dipole moment integrated over the Brillouin zone,
\begin{align}
    \overline g^{\,xx}_{nm}\equiv\int\!\!\frac{d^2\bk}{(2\pi)^2} g^{xx}_{nm}(\bk)\,.
\end{align}
Using Eq.~\eqref{eqn:gnmAppendix} and Eq.~\eqref{eqn:relationG}, the explicit results for the quantum metric in Eq.~\eqref{eq: nondegenerate qmetrics appendix} and Eq.~\eqref{eq: degenerate qmetrics appendix}, and $g^{xx}_{+-}=0$, we can explicitly calculate the quantum metrics integrated over the Brillouin zone $\overline g^{\, xx}_{n}$ and $\overline g^{\,xx}_{(nm)}$ for $n,m=-,0,+$, respectively, and obtain 
\begin{gather}
    \overline g^{\,xx}_{-}=\overline g^{\,xx}_{-0} = \frac{c}{2}\,, \hspace{1cm}
    \overline g_{0}^{\,xx}=\overline g^{\,xx}_{-0}+\overline g^{\,xx}_{+0} = c\,, \hspace{1cm}
    \overline g_{+}^{\,xx}=\overline g^{\,xx}_{+0} = \frac{c}{2}
    \label{eqn: explicit QMnondegenerate}
\end{gather}
and
\begin{gather}
    \overline g_{(+0)}^{\,xx} =\overline g^{\,xx}_{-0}= \frac{c}{2}\,,\hspace{1cm}
    \overline g_{(-0)}^{\,xx} =\overline g^{xx}_{+0}= \frac{c}{2}\,, \hspace{1cm}
    \overline g_{(+-)}^{\,xx} =\overline g^{\,xx}_{-0}+\overline g^{\,xx}_{+0}= c
    \label{eqn: explicit QMdegenerate}
\end{gather}
These results are used in the discussion of the results shown in Fig.~\ref{fig:fig2} and Fig.~\ref{fig:fig3} in the main text. By Eq.~\eqref{eqn: explicit QMnondegenerate} and Eq.~\eqref{eqn: explicit QMdegenerate}, the transition dipole moments $\overline g^{\,xx}_{nm}$ integrated over the Brillouin zone are fully determined. We read off 
\begin{gather}
    \overline g^{\,xx}_{+0}=\overline g^{\,xx}_{0+}= \frac{c}{2}\,,\hspace{1cm}
    \overline g^{\,xx}_{-0}=\overline g^{\,xx}_{0-}= \frac{c}{2}\,,\hspace{1cm}
    \overline g^{\,xx}_{+-}=\overline g^{\,xx}_{-+}= 0 \, .
\end{gather}
Thus, we obtain the explicit analytic formula of the quantum metric contribution,
\begin{align}
    \sigma^{xx,s}_\text{inter}=\frac{e^2}{h}4\pi\Big[\overline g^{xx}_{+0}\,\,w^{\text{inter},s}_{+0}+\overline g^{xx}_{-0}\,\,w^{\text{inter},s}_{-0}+\overline g^{xx}_{+-}\,\,w^{\text{inter},s}_{+-}\Big]=\frac{e^2}{h}2\pi c\Big[w^{\text{inter},s}_{+0}+w^{\text{inter},s}_{-0}\Big] \, ,
    \label{eqn: explicit formula}
\end{align}
where we used that $\overline g^{\,xx}_{nm}=\overline g^{\,xx}_{mn}$ and $h=2\pi\hbar$. The transitions are weighted by
\begin{align}
    w^{\text{inter},s}_{nm}=-\pi(\varepsilon_{n}\!-\!\varepsilon_{m})^2\mathcal{A}_n\mathcal{A}_m
    \label{eqn: weights}
\end{align}
with spectral function
\begin{align}
    \cA_n=\frac{1}{\pi}\frac{\Gamma}{\Gamma^2+\big(\varepsilon_n-\mu\big)^2}\,.
\end{align}
\begin{figure*}
    \centering
    \includegraphics[width=0.45\textwidth]{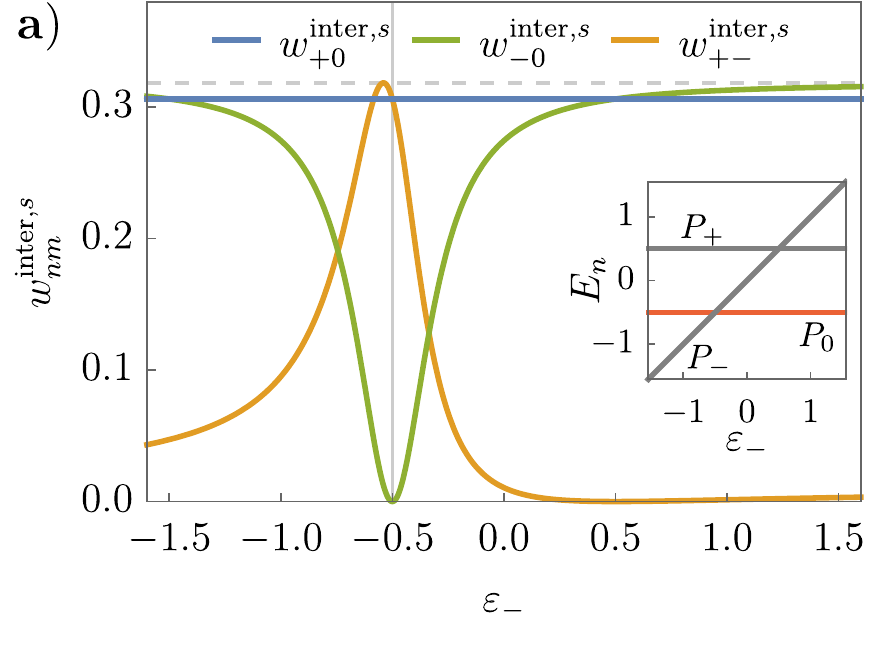}
    \includegraphics[width=0.45\textwidth]{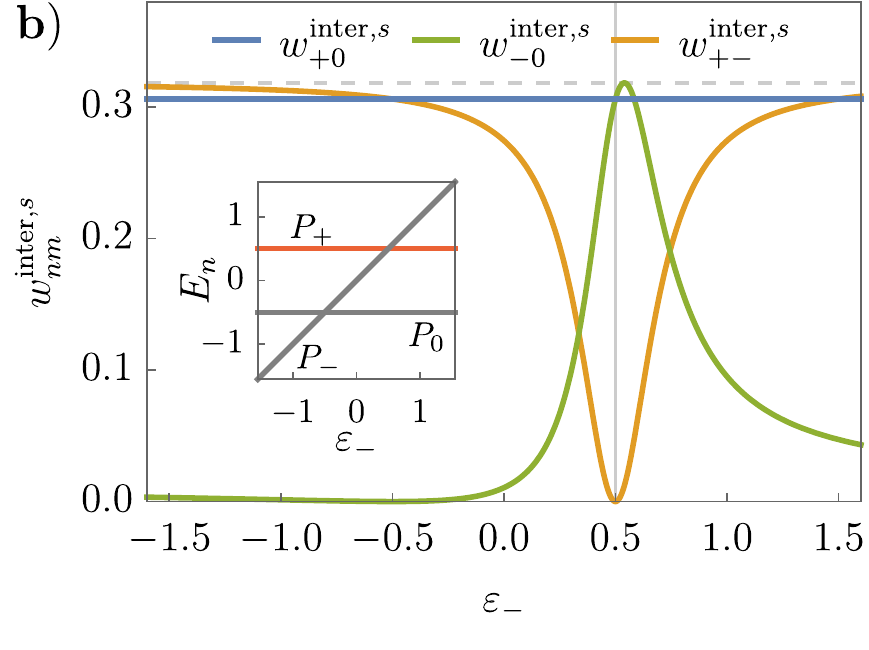}
    \includegraphics[width=0.45\textwidth]{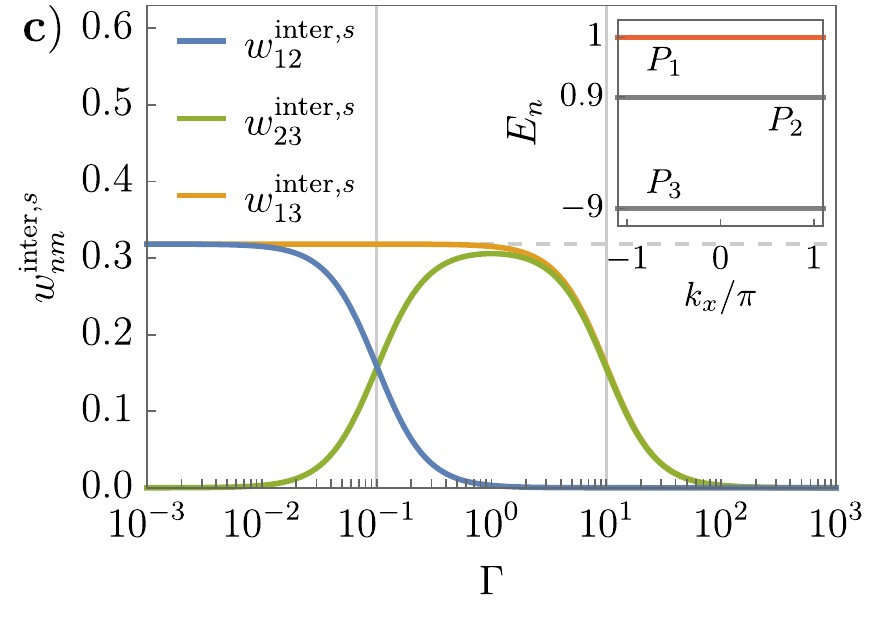}
    %\internallinenumbers
    \caption{The three $w^{\text{inter}}_{nm}$ for the corresponding settings of Fig.~\ref{fig:fig2} and Fig.~\ref{fig:fig3}. Here, we fixed $\Gamma=0.2$. The dashed line corresponds to $1/\pi$.} 
    \label{fig:figS1}
\end{figure*}
Note that $w^{\text{inter},s}_{nm}\approx 1/\pi$ if the chemical potential is fixed to one of the flat bands \cite{Mitscherling2022}. In Fig.~\ref{fig:figS1}, we plot the $w^{\text{inter},s}_{nm}$ defined in Eq.~\eqref{eqn: weights}, which completely determine the behavior of the quantum metric contribution according to Eq.~\eqref{eqn: explicit formula}. For instance, we see that the asymmetric behavior of the drop in Fig.~\ref{fig:fig2} b) can be traced back to $w^{\text{inter},s}_{-0}$, in particular, to the prefactor $(\varepsilon_--\varepsilon_0)^2$. In Fig.~\ref{fig:figS1} a) and b), we indicate the energies at which the bands are precisely degenerate by vertical gray lines. In Fig.~\ref{fig:figS1}, the vertical gray lines indicate the crossover regimes given by the band gaps. 

\subsubsection{Effect of a finite bandwidth}

We introduce a momentum dependence of the bands to the model in Eq.~\eqref{eq: Bloch Hamiltonian spin-1 model appendix} (or Eq.~\eqref{eq: Bloch Hamiltonian spin-1 model} in the main text),
\begin{align}
    \varepsilon_{n,\bk}=\varepsilon_{n}-W(\cos k_x+\cos k_y)\,.
\end{align}
Thus, the intraband contribution $\sigma^{xx}_{\text{intra}}$ given in Eq.~\eqref{eqn:sigmaIntraAppendix} is no longer zero. As discussed in detail in Ref.~\cite{Mitscherling2022}, the intraband contribution dominates for $\Gamma\lesssim W$ with $\sigma^{xx}_{\text{intra}}\propto 1/\Gamma$, whereas the quantum metric contribution $\sigma^{xx,s}_\text{inter}\propto \Gamma$ (cf., Eq.~\eqref{eqn:sigmaSclean} in the main text). For $\Gamma\gtrsim W$, the intraband contribution is suppressed and the quantum metric contribution dominates, since it is effectively flat with $|\varepsilon_{n,\bk}-\mu|\ll \Gamma$ (cf., Eq.~\eqref{eqn:sigmaFlat} in the main text). In Fig.~\ref{fig:figS2}, we show the impact of the intraband contribution onto the total conductivity $\sigma^{xx}=\sigma^{xx}_\text{intra}+\sigma^{xx,s}_\text{inter}$, which confirms these general statements. As long as the bandwidth is smaller than the band gaps, the phenomenology discussed in the main text still holds. Here, we also changed the band gap $\Delta_{13}$ between the highest and lowest band from $10$ to $5$. We see that our conclusions are unaffected, which explicitly shows that the regimes are essentially controlled by the corresponding energy scales only.
\begin{figure*}
    \centering
    \includegraphics[width=0.45\textwidth]{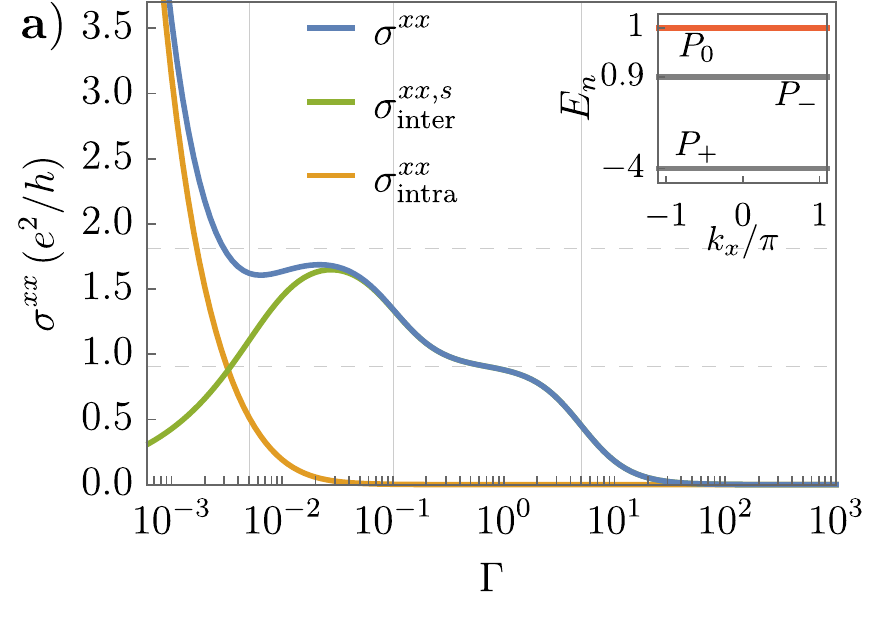}
    \includegraphics[width=0.45\textwidth]{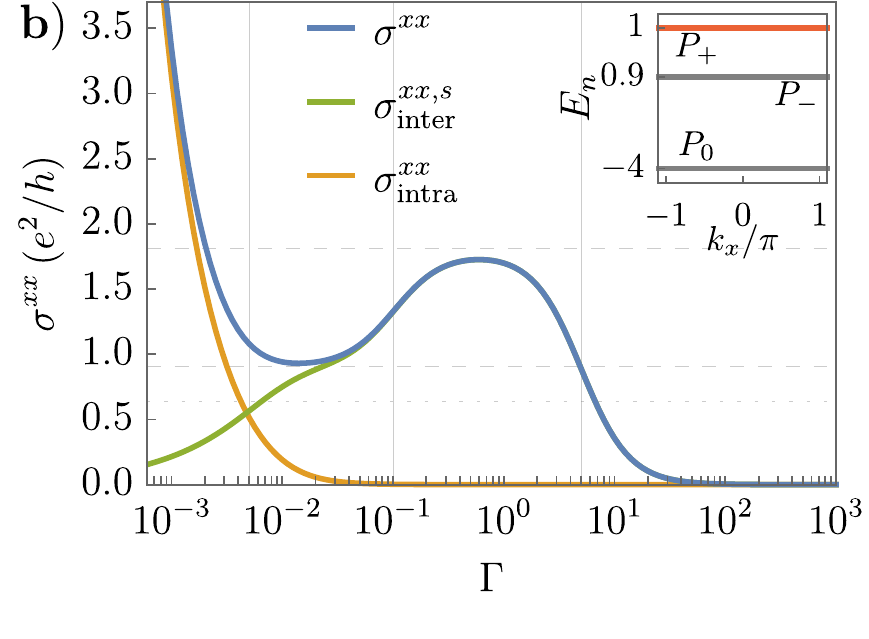}
    \includegraphics[width=0.45\textwidth]{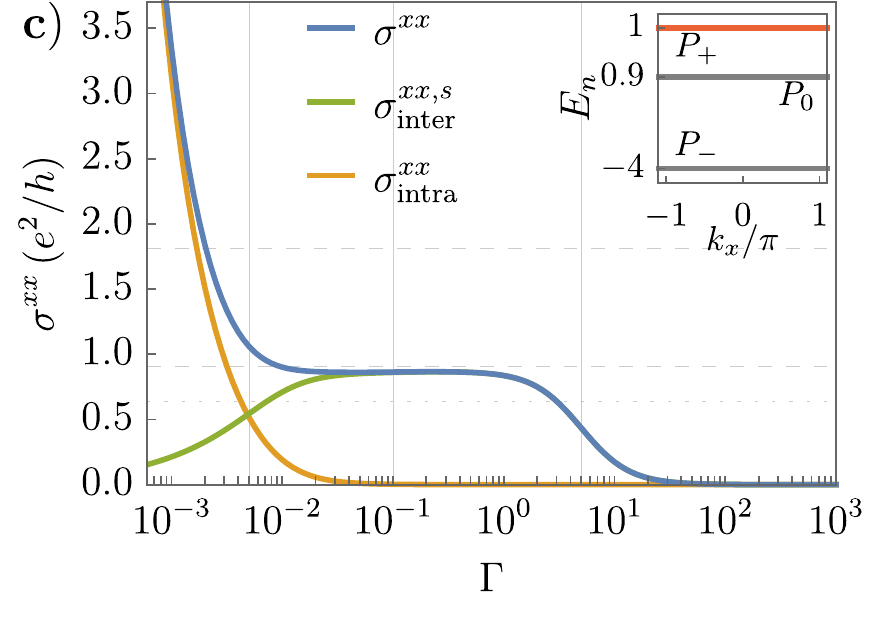}
    %\internallinenumbers
    \caption{The same plots as shown in Fig.~\ref{fig:fig3} in the main text, here with dispersive bands of bandwidth $W=0.05$ and $\Delta_{13}=5$. The dashed lines correspond to $2c$ and $4c$ given in Eq.~\eqref{eqn:valueC}. The dotted line is the lower bound due to the finite Chern number $|C_\pm|= 2$ of band $+$ and $-$ \cite{Mitscherling2022}.} 
    \label{fig:figS2}
\end{figure*}
\end{widetext}

\end{document}